# Cognitive MIMO-RF/FSO Cooperative Relay Communication with Mobile Nodes and Imperfect Channel State Information

Neeraj Varshney, *Student Member, IEEE*, Aditya K. Jagannatham, *Member, IEEE*, and Pramod K. Varshney, *Life Fellow, IEEE*

*Abstract*—This work analyzes the performance of an underlay cognitive radio based decode-and-forward mixed multiple-input multiple-output (MIMO) radio frequency/free space optical (RF/FSO) cooperative relay system with multiple mobile secondary and primary user nodes. The effect of imperfect channel state information (CSI) arising due to channel estimation error is also considered at the secondary user transmitters (SU-TXs) and relay on the power control and symbol detection processes respectively. A unique aspect of this work is that both fixed and proportional interference power constraints are employed to limit the interference at the primary user receivers (PU-RXs). Analytical results are derived to characterize the exact and asymptotic outage and bit error probabilities of the above system under practical conditions of node mobility and imperfect CSI, together with impairments of the optical channel, such as path loss, atmospheric turbulence, and pointing errors, for orthogonal space-time block coded transmission between each SU-TX and relay. Finally, simulation results are presented to yield various interesting insights into the system performance such as the benefits of a midamble versus preamble for channel estimation.

## I. INTRODUCTION

RADIO FREQUENCY (RF) spectrum scarcity is a rising concern in modern wireless cellular networks, due to the ever increasing number of users and data intensive applications in current and next generation wireless systems [1], [2]. This has led to the development of the cognitive radio paradigm, and specifically the underlay cognitive radio mode, for enabling limited spectrum access by cognitive (or secondary) users, while restricting the interference caused to the primary network. In order to limit the interference to the primary receivers (PU-RXs), the transmit power of the secondary RF transmitters (SU-TXs) in the underlay mode has to be strictly regulated. This significantly affects the reliability of cognitive radio systems, while also leading to unstable transmission and restricted coverage [3], [4]. The reliability and data rates of cognitive RF transmissions can be significantly enhanced by employing multiple-input multiple-output (MIMO) technology, through diversity and spatial multiplexing, respectively,

Neeraj Varshney and Aditya K. Jagannatham are with the Department of Electrical Engineering, Indian Institute of Technology Kanpur, Kanpur UP 208016, India (e-mail:neerajv@iitk.ac.in; adityaj@iitk.ac.in).
Pramod K. Varshney is with the Department of Electrical Engineering & Computer Science, Syracuse University, Syracuse, NY 13244, USA (e-mail:varshney@syr.edu).

for the same transmit power [5]. Moreover, relaying using a single broadband free space optical (FSO) link is a cost effective approach for increasing the cognitive communication range in the presence of a primary network, while supporting high speed data transmission and not causing any additional interference to primary communication over the RF bands. It is important to note that since the relay communicates both with the donor cell and cognitive terminals served by the relay, interference between the access and backhaul links must be avoided [6, Fig. 18.2]. Therefore, FSO technology for the backhaul links is an ideal choice to avoid interference with simultaneous transmissions in the cognitive radio network as shown in Fig. 1. Further, multiple existing cognitive RF devices/ users can be simultaneously connected to the network backbone via a donor cell or serving base-station (eNodeB) over a single broadband FSO link for last-mile access [7]. In addition, since broadband FSO links enable high data rate transmission over unlicensed optical bands, they also address the problem of RF spectrum scarcity in the licensed bands [2]. Hence, FSO technology is well poised for practical deployment in future wireless networks towards relieving the intense pressure on the existing RF spectral bands, while providing additional benefits such as the ability to be deployed rapidly, low costs, security, and resilience to RF interference. Due to the above advantages of FSO technology in cognitive radio network scenarios, this work, therefore, considers a cognitive MIMO-RF/FSO cooperative relay communication system. Orthogonal space-time block coded (OSTBC) MIMO transmission is well suited for practical implementation in such systems due to the associated low complexity of optimal maximum-likelihood decoding [8] while also not requiring any channel knowledge of the SU-relay links at the SU-TX nodes. However, mobility of the SU-TXs and PU-RXs induces Doppler [9], which leads to time-selective fading in the RF links and the ensuing degradation of the performance of the system. Moreover, in time varying scenarios, it is also difficult to obtain perfect channel state information (CSI) of the cross channel link at the SU-TX and cognitive link at the relay for power control and symbol detection respectively. Hence, end-to-end performance analysis of underlay cognitive mixed MIMO-OSTBC RF/FSO cooperative communication systems, considering also artifacts such as time-selective RF links arising due to node mobility together with imperfect CSI from channel estimation error, is critically important from a practical perspective and is, therefore, one of the central aims of the work presented in this paper. Next, we present a brief overview and comprehensive survey of related works in the existing literature and also clarify our contributions.

### A. Related work

The work in [10] first introduced an asymmetric amplify-and-forward (AF) dual-hop RF/FSO system, where the RF link between the source and relay, and the FSO link between the relay and destination are assumed to be Rayleigh and Gamma-Gamma faded respectively. The authors in [11] and [12] extended the work to include misalignment fading (pointing errors) in the FSO link for intensity modulation/direct detection (IM/DD) and heterodyne detection schemes, respectively. Performance analysis of a dual-hop AF RF/FSO system with Rayleigh faded RF channel and generalized $M$-distribution faded optical channel is presented in [13]. The effects of fading, turbulence, and pointing error on the outage probability, average bit error rate (BER), and channel capacity of an asymmetric AF RF/FSO system considering a Nakagami-$m$ distributed RF link and Gamma-Gamma distributed FSO link with pointing errors have been analyzed in [14]. The analysis therein has been further extended to include the effect of outdated channel state information (CSI) at the relay in [15]. In contrast to AF relaying considered in works such as [10]–[15], DF relaying based mixed RF/FSO systems have been analyzed in [7], [16]–[20]. However, none of the existing works above consider the cognitive radio paradigm to alleviate the problem of spectrum scarcity.

Towards alleviating spectrum scarcity, the authors in [1], [2], [21] proposed and analyzed the performance of a dual-hop communication system composed of asymmetric RF and FSO links for underlay cognitive networks. In [1], [2], the presence of multiple secondary users is assumed for transmission to the secondary base station over Rayleigh faded RF links. The work in [21] investigated the performance of a system with a single SU-TX with independent and non-identically distributed Nakagami-$m$ fading RF links. The outage performance of a multi-user mixed underlay RF/FSO network with multiple destination nodes has been analyzed in [22]. However, studies in works [1], [2], [21], [22] employ an amplify-and-forward (AF) protocol at the relay that is not easily realizable in practice since both the RF and FSO links operate at different frequencies. The relay therefore has to decode the information symbol prior to modulation over an optical carrier. In order to overcome the above challenge, our previous work in [23] proposed and analyzed the outage performance of a DF based underlay cognitive mixed MIMO-RF/FSO cooperative system with single SU-TX and PU-RX nodes. However, none of the works reviewed above consider the effect of practical degradations such as time-selective fading, imperfect channel knowledge at the SU-TX and relay nodes to present a comprehensive analysis that yields various insights into the end-to-end performance of a practical system.

### B. Contributions

The novel contributions of the paper including the key technical differences with respect to previous works are summarized below.

- This work analyzes the end-to-end performance of a multi-user underlay cognitive mixed MIMO-OSTBC RF/FSO DF cooperative system where the RF links between the SU-TX and relay nodes, and SU-TX and PU-RX nodes experience time-selective fading, while the FSO link is affected by optical channel impairments such as path loss, atmospheric turbulence, and pointing errors. Exact and asymptotic results are derived for the outage and bit-error probabilities incorporating also mobility of SU-TXs and PU-RXs with arbitrary speeds. Further, the effect of imperfect channel state information is also considered towards power control and symbol detection at each SU-TX and relay respectively. For symbol detection, only imperfect channel knowledge is considered at the relay, which is estimated only once at the $L$th signaling instant of each frame. For power control in the underlay mode, each SU-TX obtains the cross channel knowledge only in the beginning of each frame. Further, employing the Kalman approach [24], the available estimate is subsequently used to predict the cross channel gains at each codeword transmission in the corresponding frame. In contrast, the previous work in [23] analyzed the outage performance considering only a restricted scenario with single static SU-TX and PU-RX nodes and perfect CSI. Moreover, results were not presented therein for the resulting BER performance.
- Additionally, a novel composite power constraint, comprising of both fixed and proportional interference power constraints, has been employed to limit the interference at the PU-RXs. This plays an important role in limiting the transmit power of each SU-TX from rising excessively due to inaccurate estimates of the cross-channel gain. However, the previous work in [23] employed only the fixed power constraint. Hence, the derivations of the cumulative distribution function (CDF) and probability density function (PDF) of the SU TX-relay link SNRs considering both the fixed and proportional interference power constraints and the resulting analysis is fundamentally different in comparison to the previous works.
- Another key contribution is that the current work presents a general framework to model time-selective fading in the SU-TX and relay links that can accommodate pilots at any arbitrary location in the block for channel estimation through appropriate choice of the parameter $L$. It is then shown that the midamble results in a significant decrease in the outage probability of the cognitive system in comparison to the preamble.
- Simulation results explicitly demonstrate that the system performance degrades drastically for the scenario when only the SU-TX nodes are mobile as compared to one in which each of the SU-TX and PU-RX nodes are static. On the other hand, interestingly, the end-to-end performance can be seen to be better for the scenario with only the PU-RX nodes mobile in comparison to the scenarios with either the SU-TXs are mobile or each SU-TX and PU-RX is static. Also, the outage and error probabilities in the low and moderate SNR regimes further decrease with increasing speed of the PU-RX nodes. Moreover, the current system with fixed and proportional interference power constraints has an improved primary network performance in comparison to the previous scheme in [23], which considers only the fixed interference power constraint for the secondary user transmissions, while neglecting the proportional interference power constraint.

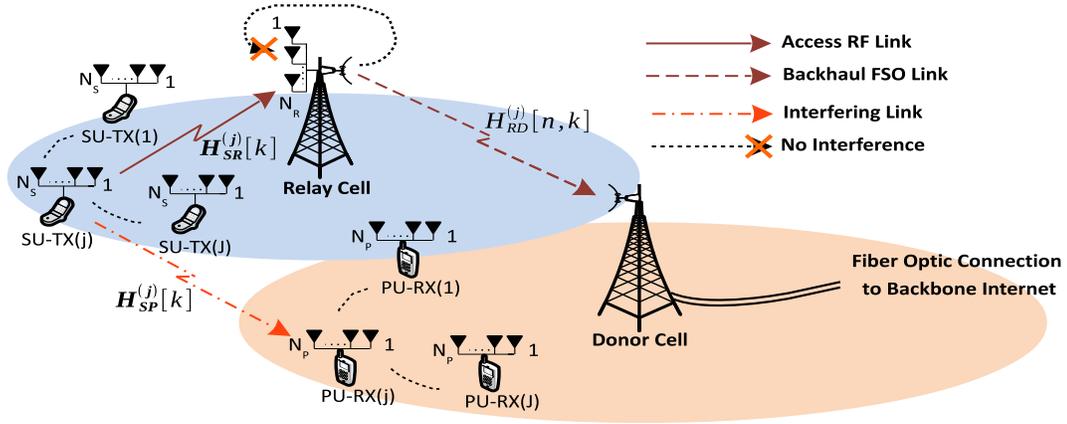

Fig. 1. Schematic diagram of the multi-user cognitive MIMO-RF/FSO DF cooperative system where SU-TX($j$) and PU-RX($j$) denote the $j$th secondary user transmitter and $j$th primary user receiver respectively.

In this work, the time variation of the wireless fading channel is captured by a first order autoregressive (AR1) Jakes model. This model best captures the wireless channel for practical scenarios with mobile nodes, as described in the work on vehicular communication [25] for wireless access in vehicular environments (WAVE) related studies. Also, this has been well researched for mobile-to-mobile communication scenarios in [26] and channel models based on the Jakes Doppler power spectrum have been proposed in [27]. The widespread popularity and appeal of this model is demonstrated by its use in several related existing works on cooperative communication with mobile nodes such as [28]–[32]. Further, the work in [33] has verified the applicability of the AR1 model, establishing its theoretical accuracy. Thus, the AR1 Jakes model is best suited to capture the behavior of the time-selective channel towards performance analysis in wireless scenarios with node mobility.

*C. Notations and Organization*

The complete list of notations and mathematical functions used in this work is given in Table I. The rest of the paper is organized as follows. The system model for the cognitive network with SU-TX and PU-RX nodes mobility, and imperfect channel estimates is presented in Section II. Exact and asymptotic results are derived for the outage and bit error probabilities in Section III and Section IV respectively. Simulation results under various node mobility scenarios are presented in Section V, followed by conclusions in Section VI.

TABLE I
List of notations and mathematical functions

| Quantity | Description |
|---|---|
| $N_S$ | Number of transmit antennas at each SU-TX |
| $N_P$ | Number of receive antennas at each PU-RX |
| $N_R$ | Number of receive antennas at relay |
| $R_c$ | Rate of the OSTBC |
| $N_b$ | Number of OSTBC codeword matrices in a frame |
| $N_a$ | Number of nonzero symbol transmissions per codeword instant |
| $\nu_{SR}^{(j)}$ | Speed of $j$th mobile SU-TX |
| $\nu_{SP}^{(j)}$ | Speed of $j$th mobile PU-RX |
| $P_{S,k}^{(j)}$ | $j$th SU-TX transmit power |
| $\Re$ | Photodetector responsivity |
| $A$ | Photodetector area |
| $q$ | Standard electronic charge |
| $\triangle f$ | Noise equivalent bandwidth of the receiver |
| $I_b$ | Background light irradiance |
| $k_b$ | Boltzmann's constant |
| $T_k$ | Temperature in Kelvin |
| $\zeta$ | Modulation index |
| $F_n$ | Thermal noise enhancement factor due to amplifier noise |
| $R_L$ | Load resistance |
| $\mu_\theta$ | Average electrical SNR |
| $\alpha$ | Large-scale scintillation parameter |
| $\beta$ | Small-scale scintillation parameter |
| $w_e$ | Equivalent beam-width radius |
| $\sigma_s$ | Standard deviation of the pointing error displacement at the receiver |
| $P_A$ | Acceptable interference power at PU-RXs |
| $P_M^{(j)}$ | Peak transmit power at $j$th SU-TX |
| $\rho_{SP,j}$ | Correlation parameter for the $j$th SU TX-$j$th PU RX link |
| $\rho_{SR,j}$ | Correlation parameter for the $j$th SU TX-relay link |
| $\gamma_{\text{th}}$ | SNR outage threshold |
| $G_{p,q}^{l,m}(x\mid {a_1,...,a_p \atop b_1,...,b_q})$ | Meijer's G-function |
| $\Gamma(x)$ | Gamma function |
| $\gamma(s,x)$ | Lower incomplete Gamma function |
| $\Gamma(s,x)$ | Upper incomplete Gamma function |
| $G_{-,-:-,-:-,-}^{-,-:-,-:-,-}(-)$ | Generalized Meijer's G-function of two variables |

## II. COGNITIVE SYSTEM MODEL

Consider a multi-user underlay cognitive mixed MIMO-RF/FSO DF cooperative relay network as shown in Fig. 1, where multiple SU-TXs transmit the information symbols at the same time using the orthogonal spectral resources of the primary network. Each SU-TX employs $N_S$ transmit antennas, and each of the PU-RX nodes and relay have $N_P$ and $N_R$ receive antennas respectively. In this system, end-to-end cognitive communication with multiple SU-TXs is carried out in two phases. In the first phase, each of the mobile SU-TXs communicate with the relay where the $j$th SU-TX moving at a speed of $\nu_{SR}^{(j)}$, transmits an OSTBC block $\mathbf{X}_{SR}^{(j)}[k] \in \mathbb{C}^{N_S \times T}$ of rate $R_c = \frac{B}{T}$ to the relay node using the spectral resource of the $j$th primary user, where a block encodes $B$ symbols, $T$ denotes the block length, and $k$ represents the $k$th coded block in a frame of $N_b$ codeword matrices. The multi-antenna $j$th PU-RX is mobile with speed $\nu_{SP}^{(j)}$. The received codeword $\mathbf{Y}_{SR}^{(j)}[k] \in \mathbb{C}^{N_R \times T}$ at the relay corresponding to the transmission by the $j$th SU-TX is given as

$$\mathbf{Y}_{SR}^{(j)}[k] = \sqrt{\frac{P_{S,k}^{(j)}}{R_c N_S}} \mathbf{H}_{SR}^{(j)}[k] \mathbf{X}_{SR}^{(j)}[k] + \mathbf{W}_{SR}^{(j)}[k], \quad (1)$$

where $P_{S,k}^{(j)}$ is the $j$th SU-TX transmit power and $\mathbf{H}_{SR}^{(j)}[k] \in \mathbb{C}^{N_R \times N_S}$ denotes the Rayleigh[1] fading MIMO channel matrix between the $j$th SU-TX and the relay. Similar to [36], the additive white Gaussian noise plus primary user interference matrix $\mathbf{W}_{SR}^{(j)}[k] \in \mathbb{C}^{N_R \times T}$ comprises of entries that are modeled as independent and identically distributed (i.i.d.) complex Gaussian random variables with mean zero and variance $\eta_0$. In the second phase, the relay retransmits the decoded symbols of each SU-TX to the desired eNodeB over an FSO link after modulating over an optical carrier. The optical channel for the FSO link corresponding to the $j$th SU-TX transmission of the $n$th symbol of the $k$th coded block can be modeled as [37, Eq. (2)]

$$H_{RD}^{(j)}[n,k] = H_l^{(j)}[n,k] H_p^{(j)}[n,k] H_a^{(j)}[n,k], \quad (2)$$

where $n \in \{1, 2, \cdots, B\}$, $H_l^{(j)}[n,k]$ is the deterministic path loss, $H_p^{(j)}[n,k]$ represents the misalignment fading due to pointing errors, and $H_a^{(j)}[n,k]$ represents the atmospheric turbulence-induced fading modeled using the Gamma-Gamma distribution.

Let the quantities $\Re$, $A$, $q$, and $\triangle f$ denote the photodetector responsivity, photodetector area, standard electronic charge, and the noise equivalent bandwidth of the receiver respectively. For coherent heterodyne detection, the instantaneous SNR of the FSO link is given from [38, Eq. (2)] as $\gamma_{RD}^{(j)}[n,k] \approx \frac{\Re A}{q \triangle f} H_{RD}^{(j)}[n,k]$. On the other hand, for intensity modulation/direct detection (IM/DD) at the eNodeB, the instantaneous SNR of the FSO link is $\gamma_{RD}^{(j)}[n,k] = \frac{(\Re A \zeta)^2}{2\triangle f(q\Re A I_b + 2k_b T_k F_n/R_L)} \left(H_{RD}^{(j)}[n,k]\right)^2$ [38, Eq. (4)], where $I_b$ denotes the background light irradiance, $k_b$ represents the Boltzmann's constant, $T_k$, $\zeta$, $F_n$, $R_L$ denote the temperature in Kelvin, modulation index, thermal noise enhancement factor due to amplifier noise, and the load resistance respectively. The PDF and CDF of $\gamma_{RD}^{(j)}[n,k]$ in terms of Meijer's G-function[2] $G_{p,q}^{l,m}\left(x \;\middle|\; \begin{smallmatrix} a_1,\ldots,a_p \\ b_1,\ldots,b_q \end{smallmatrix}\right)$ [39] are given as [40, Eqs. (4), (5)]

$$f_{\gamma_{RD}^{(j)}[n,k]}(x)$$
$$= \frac{\xi^2}{\theta x \Gamma(\alpha)\Gamma(\beta)} G_{1,3}^{3,0}\left(\alpha\beta\left(\frac{x}{\mu_\theta}\right)^{1/\theta} \;\middle|\; \begin{smallmatrix} \xi^2 + 1 \\ \xi^2, \alpha, \beta \end{smallmatrix}\right), \quad (3)$$

$$F_{\gamma_{RD}^{(j)}[n,k]}(x) = \Theta_1 G_{\theta+1,3\theta+1}^{3\theta,1}\left(\frac{\Theta_2}{\mu_\theta} x \;\middle|\; \begin{smallmatrix} 1, \Theta_3 \\ \Theta_4, 0 \end{smallmatrix}\right), \quad (4)$$

where $\theta = 1$ for heterodyne detection and $\theta = 2$ for IM/DD, and $\mu_\theta$ is the average electrical SNR [40]. The large-scale and small-scale scintillation parameters respectively are $\alpha$ and $\beta$, and $\xi = w_e/2\sigma_s$, where $w_e$ is the equivalent beam-width radius and $\sigma_s$ is the standard deviation of the pointing error displacement at the receiver. The relevant quantities in (4) are $\Theta_1 = \frac{\theta^{\alpha+\beta-2}\xi^2}{(2\pi)^{\theta-1}\Gamma(\alpha)\Gamma(\beta)}$, $\Theta_2 = \frac{(\alpha\beta)^\theta}{\theta^{2\theta}}$, $\Theta_3 = \frac{\xi^2+1}{\theta}, \ldots, \frac{\xi^2+\theta}{\theta}$ is a group of $\theta$ terms, and $\Theta_4 = \frac{\xi^2}{\theta}, \ldots, \frac{\xi^2+\theta-1}{\theta}, \frac{\alpha}{\theta}, \ldots, \frac{\alpha+\theta-1}{\theta}, \frac{\beta}{\theta}, \ldots, \frac{\beta+\theta-1}{\theta}$ is a group of $3\theta$ terms.

It is important to note that the transmission over the FSO link in the second phase does not create any interference to the PU-RXs owing to the highly directional nature of the optical beam. However, to prevent the interference at the $j$th PU-RX from exceeding an acceptable threshold $P_A$ in the first phase, the $j$th SU-TX transmit power must satisfy the proportional and fixed interference power constraints, similar to works [3], [4], [41], as given below

$$P_{S,k}^{(j)} = \begin{cases} P_M^{(j)} & \text{if } \|\widehat{\mathbf{H}}_{SP}^{(j)}[k]\|_F^2 \leq \frac{P_A}{P_M^{(j)}}, \\ \dfrac{P_A}{\|\widehat{\mathbf{H}}_{SP}^{(j)}[k]\|_F^2} & \text{if } \|\widehat{\mathbf{H}}_{SP}^{(j)}[k]\|_F^2 > \frac{P_A}{P_M^{(j)}}. \end{cases} \quad (5)$$

The quantity $P_M^{(j)}$ denotes the maximum transmit power available at the $j$th SU-TX, while $\|\widehat{\mathbf{H}}_{SP}^{(j)}[k]\|_F^2$ denotes the estimate of the $j$th SU TX-$j$th PU RX channel gain available at the $j$th SU-TX. Note that in the proportional interference power constraint, the interference power at the $j$th PU-RX is proportional to the maximum transmit power $P_M^{(j)}$ [3]. On the other hand, under the fixed interference power constraint, the peak interference power $P_A$ at the $j$th PU-RX is fixed and independent of the maximum transmit power $P_M^{(j)}$ available at the $j$th SU-TX [4]. In practice, it is difficult to obtain the knowledge of the cross channel $\widehat{\mathbf{H}}_{SP}^{(j)}[k]$ at each signaling instant $k = 1, 2, \cdots, N_b$ due to the time-varying nature of the $j$th PU RX-$j$th SU TX link. This work, therefore, considers that the $j$th SU-TX estimates the cross channel $\mathbf{H}_{SP}^{(j)}[1] \in \mathbb{C}^{N_P \times N_S}$ only at the beginning of each frame, with the estimate $\widehat{\mathbf{H}}_{SP}^{(j)}[1] = \mathbf{H}_{SP}^{(j)}[1] + \mathbf{H}_{\epsilon,SP}^{(j)}[1]$, where $\mathbf{H}_{\epsilon,SP}^{(j)}[1] \in \mathbb{C}^{N_P \times N_S}$ represents the estimation error. The

---

[1] As stated in [34], the links between the SU-TX and PU-RX, SU-TX and relay can be assumed to be independent Rayleigh fading for the scenarios when SU-TX is far away from the PU-RX and relay. Further, note that for Nakagami-$m$ fading links, the expressions for the CDF and the PDF of the resulting instantaneous received SNR in (12) with imperfect channel estimates are analytically intractable, since the entries of the effective time-selective channel matrices $\widehat{\mathbf{H}}_{SR}^{(j)}[L]$ and $\widehat{\mathbf{H}}_{SP}^{(j)}[k]$ are distributed as the sum of Nakagami-$m$ and circularly symmetric complex Gaussian random variables. Therefore, closed-form expressions for the outage and error probabilities considering time-selective Nakagami-$m$ fading SU TX-PU RX and SU TX-relay links with imperfect channel estimates cannot be easily obtained. However, a complete analysis for the outage probability over time-selective Nakagami-$m$ fading links with perfect channel estimates and the results is given in the technical report [35]. These results are not included here for brevity.

[2] The Meijer's G-function can be readily evaluated using MATLAB and the corresponding command for evaluating $G_{p,q}^{l,m}\left(x \;\middle|\; \begin{smallmatrix} a_1,\ldots,a_p \\ b_1,\ldots,b_q \end{smallmatrix}\right)$ with $l = 3, m = 1, p = 2$, and $q = 4$ is given as, double(evalin(symengine, sprintf('meijerG(3, 1,[%d,%d], [%d,%d,%d,%d],%d)', $a_1, a_2, b_1, b_2, b_3, b_4, x$))).

entries of $\mathbf{H}_{SP}^{(j)}[1]$ and $\mathbf{H}_{\epsilon,SP}^{(j)}[1]$ are modeled as i.i.d. complex Gaussian random variables with variances $\delta_{SP,j}^2$ and $\sigma_{\epsilon_{SP,j}}^2$ respectively. Imperfect channel knowledge at the $j$th SU-TX can be obtained through limited feedback from the $j$th PU-RX similar to works such as [42], [43]. Further, similar to the Kalman approach, the available channel knowledge $\widehat{\mathbf{H}}_{SP}^{(j)}[1]$ at the $j$th SU-TX is subsequently used to obtain the cross channel gains at $k = 1, 2, \cdots, N_b$ signaling instants as, $G_{SP,k}^{(j)} = \|\widehat{\mathbf{H}}_{SP}^{(j)}[k]\|_F^2 = \rho_{SP,j}^{2(k-1)} \|\widehat{\mathbf{H}}_{SP}^{(j)}[1]\|_F^2$, where $\rho_{SP,j}$ denotes the correlation parameter for the $j$th SU TX-$j$th PU RX link. Also, as derived in detail in [9], the correlation coefficient $\rho$ for the Jakes model is derived by computing the expected correlation over a large number of multipath components with angles of arrival uniformly distributed in $[-\pi, \pi]$. Thus, it also inherently captures the direction of arrival of the mobile users, as justified by [33], and is employed by several related works such as [32] and the references therein.

Moreover, the MIMO channel between the $j$th SU-TX and relay also experiences time-selective fading due to the mobile nature of the $j$th SU-TX. Therefore, it is often difficult to obtain the instantaneous CSI corresponding to each transmitted codeword matrix $\mathbf{X}_{SR}^{(j)}[k], 1 \leq k \leq N_b$ at the relay for symbol detection. Consequently, it is assumed that the relay node has only imperfect knowledge of the channel matrix $\widehat{\mathbf{H}}_{SR}^{(j)}[L] = \mathbf{H}_{SR}^{(j)}[L] + \mathbf{H}_{\epsilon,SR}^{(j)}[L]$, estimated once at the $L$th signaling instant of each frame and is subsequently employed to detect each codeword $\mathbf{X}_{SR}^{(j)}[k], 1 \leq k \leq N_b$ in the corresponding frame. The entries of the true channel matrix $\mathbf{H}_{SR}^{(j)}[L]$ and the error matrix $\mathbf{H}_{\epsilon,SR}^{(j)}[L]$ are modeled as i.i.d. complex Gaussian random variables with variances $\delta_{SR,j}^2$ and $\sigma_{\epsilon_{SR,j}}^2$ respectively. It is important to mention that the above assumption is a valid one since the receiver tracking loop in a typical wireless system cannot estimate the MIMO channel in each signaling period $k$ due to the significantly higher pilot overhead such a scheme would incur. Also, note that in contrast to the works in [32], [44] and the references therein, the proposed model is more general and can accommodate both a preamble or midamble for channel estimation through appropriate choice of the parameter $L$. The basic AR1 process to model a time-selective fading SISO wireless channel is given in [32, Eq. (4)]. Using this, the time-selective fading model for the $j$th SU TX-relay MIMO link at time instant $k$ can be determined as

$$\mathbf{H}_{SR}^{(j)}[k] = \begin{cases} \rho_{SR,j}\mathbf{H}_{SR}^{(j)}[k+1] \\ \quad + \sqrt{1-\rho_{SR,j}^2}\mathbf{E}_{SR}^{(j)}[k], & \text{if } 1 \leq k < L, \\ \rho_{SR,j}\mathbf{H}_{SR}^{(j)}[k-1] \\ \quad + \sqrt{1-\rho_{SR,j}^2}\mathbf{E}_{SR}^{(j)}[k], & \text{if } L < k \leq N_b, \end{cases} \quad (6)$$

$$= \begin{cases} \rho_{SR,j}^{L-k}\mathbf{H}_{SR}^{(j)}[L] + \sqrt{1-\rho_{SR,j}^2} \\ \quad \times \sum_{n=k}^{L-1} \rho_{SR,j}^{n-k}\mathbf{E}_{SR}^{(j)}[n], & \text{if } 1 \leq k < L, \\ \rho_{SR,j}^{k-L}\mathbf{H}_{SR}^{(j)}[L] + \sqrt{1-\rho_{SR,j}^2} \\ \quad \times \sum_{n=L+1}^{k} \rho_{SR,j}^{k-n}\mathbf{E}_{SR}^{(j)}[n], & \text{if } L < k \leq N_b, \end{cases} \quad (7)$$

where $\rho_{SR,j}$ is the correlation parameter for the $j$th SU TX-relay link and the entries of $\mathbf{E}_{SR}^{(j)}[n] \in \mathbb{C}^{N_R \times N_S}$ are modeled as circularly symmetric complex Gaussian random variables with variance $\sigma_{e_{SR,j}}^2$. Employing now $\mathbf{H}_{SR}^{(j)}[L] = \widehat{\mathbf{H}}_{SR}^{(j)}[L] - \mathbf{H}_{\epsilon,SR}^{(j)}[L]$ in (7) and substituting the resulting expression for $\mathbf{H}_{SR}^{(j)}[k]$ in (1), the received codeword matrix can be determined as

$$\mathbf{Y}_{SR}^{(j)}[k] = \begin{cases} \sqrt{\frac{P_{S,k}^{(j)}}{R_c N_S}}\rho_{SR,j}^{L-k}\widehat{\mathbf{H}}_{SR}^{(j)}[L]\mathbf{X}_{SR}^{(j)}[k] \\ \quad + \widehat{\mathbf{W}}_{SR}^{(j)}[k], & \text{if } 1 \leq k < L, \\ \sqrt{\frac{P_{S,k}^{(j)}}{R_c N_S}}\rho_{SR,j}^{k-L}\widehat{\mathbf{H}}_{SR}^{(j)}[L]\mathbf{X}_{SR}^{(j)}[k] \\ \quad + \widetilde{\mathbf{W}}_{SR}^{(j)}[k], & \text{if } L \leq k \leq N_b, \end{cases} \quad (8)$$

where the effective noise matrices $\widehat{\mathbf{W}}_{SR}^{(j)}[k]$ and $\widetilde{\mathbf{W}}_{SR}^{(j)}[k]$ are given as

$$\widehat{\mathbf{W}}_{SR}^{(j)}[k] = \mathbf{W}_{SR}^{(j)}[k] - \sqrt{\frac{P_{S,k}^{(j)}}{R_c N_S}}\rho_{SR,j}^{L-k}\mathbf{H}_{\epsilon,SR}^{(j)}[L]\mathbf{X}_{SR}^{(j)}[k]$$
$$+ \sqrt{\frac{P_{S,k}^{(j)}(1-\rho_{SR,j}^2)}{R_c N_S}} \sum_{n=k}^{L-1} \rho_{SR,j}^{n-k}\mathbf{E}_{SR}^{(j)}[n]\mathbf{X}_{SR}^{(j)}[k], \quad (9)$$

$$\widetilde{\mathbf{W}}_{SR}^{(j)}[k] = \mathbf{W}_{SR}^{(j)}[k] - \sqrt{\frac{P_{S,k}^{(j)}}{R_c N_S}}\rho_{SR,j}^{k-L}\mathbf{H}_{\epsilon,SR}^{(j)}[L]\mathbf{X}_{SR}^{(j)}[k]$$
$$+ \sqrt{\frac{P_{S,k}^{(j)}(1-\rho_{SR,j}^2)}{R_c N_S}} \sum_{n=L+1}^{k} \rho_{SR,j}^{k-n}\mathbf{E}_{SR}^{(j)}[n]\mathbf{X}_{SR}^{(j)}[k]. \quad (10)$$

### III. OUTAGE PROBABILITY ANALYSIS

Using (8), employing an analysis similar to that for OSTBC with fixed nodes in [45, Eq. (3)], the effective instantaneous SNR for the $n$th symbol of the $k$th codeword matrix at the relay can be obtained as

$$\gamma_{SR}^{(j)}[n,k] = \frac{P_{S,k}^{(j)}\rho_{SR,j}^{2i}\|\widehat{\mathbf{H}}_{SR}^{(j)}[L]\|_F^2}{R_c N_S \eta_{0,k}^{(j)}}, \quad (11)$$

where $i = L-k$ for $1 \leq k < L$ and $i = k-L$ for $L \leq k \leq N_b$. The effective noise power $\eta_{0,k}^{(j)}$ is given as,

$$\eta_{0,k}^{(j)} = \eta_0 + \frac{P_{S,k}^{(j)}\rho_{SR,j}^{2i}}{R_c N_S}\tilde{\sigma}_{\epsilon_{SR,j}}^2 + \frac{P_{S,k}^{(j)}(1-\rho_{SR,j}^{2i})}{R_c N_S}\tilde{\sigma}_{e_{SR,j}}^2,$$

where the quantities $\tilde{\sigma}_{\epsilon_{SR,j}}^2$ and $\tilde{\sigma}_{e_{SR,j}}^2$ are defined as, $N_a\sigma_{\epsilon_{SR,j}}^2$ and $N_a\sigma_{e_{SR,j}}^2$ respectively. The parameter $N_a$ denotes the number of nonzero symbol transmissions per codeword instant. For example, $N_a = N_S = 2$ for the Alamouti OSTBC and $N_a = N_S = 3$ for the rate $R_c = \frac{1}{2}$ code $\mathcal{G}_c^3$ [8]. Further, considering the power constraints given in (5), the above expression can be simplified as

$$\gamma_{SR}^{(j)}[n,k] = \begin{cases} \alpha_{1,k}^{(j)}\|\widehat{\mathbf{H}}_{SR}^{(j)}[L]\|_F^2 & \text{if } \|\widehat{\mathbf{H}}_{SP}^{(j)}[k]\|_F^2 \leq \frac{P_A}{P_M^{(j)}}, \\ \frac{\alpha_{2,k}^{(j)}\|\widehat{\mathbf{H}}_{SR}^{(j)}[L]\|_F^2}{\|\widehat{\mathbf{H}}_{SP}^{(j)}[k]\|_F^2 + \alpha_{3,k}^{(j)}} & \text{if } \|\widehat{\mathbf{H}}_{SP}^{(j)}[k]\|_F^2 > \frac{P_A}{P_M^{(j)}}, \end{cases} \quad (12)$$

where the various quantities are defined as $\alpha_{1,k}^{(j)} = \frac{P_M^{(j)} \rho_{SR,j}^{2i}}{R_c N_S \eta_0 + P_M^{(j)} \rho_{SR,j}^{2i} \tilde{\sigma}_{\epsilon_{SR},j}^2 + P_M^{(j)}(1-\rho_{SR,j}^{2i})\tilde{\sigma}_{e_{SR},j}^2}$, $\alpha_{2,k}^{(j)} = \frac{P_A \rho_{SR,j}^{2i}}{R_c N_S \eta_0}$, and $\alpha_{3,k}^{(j)} = \frac{P_A \rho_{SR,j}^{2i} \tilde{\sigma}_{\epsilon_{SR},j}^2 + P_A(1-\rho_{SR,j}^{2i})\tilde{\sigma}_{e_{SR},j}^2}{R_c N_S \eta_0}$. Result below gives the cumulative distribution function (CDF) and probability density function (PDF) of $\gamma_{SR}^{(j)}[n,k]$.

*Theorem 1:* The CDF $F_{\gamma_{SR}^{(j)}[n,k]}(x)$ and PDF $f_{\gamma_{SR}^{(j)}[n,k]}(x)$ of the SNR $\gamma_{SR}^{(j)}[n,k]$ for the $j$th SU TX-relay link are given as

$$F_{\gamma_{SR}^{(j)}[n,k]}(x)$$
$$= \frac{1}{\Gamma(\tau_2)} \left[ \frac{1}{\Gamma(\tau_1)} \gamma\left(\tau_1, \frac{x}{\alpha_{1,k}^{(j)} \tilde{\delta}_{SR,j}^2}\right) \gamma\left(\tau_2, \frac{P_A}{\tilde{\delta}_{SP,j,k}^2 P_M^{(j)}}\right) \right.$$
$$+ \frac{1}{(\tilde{\delta}_{SP,j,k}^2)^{\tau_2}} \left\{ \mathcal{J}\left(\tau_2; \frac{1}{\tilde{\delta}_{SP,j,k}^2}\right) - \exp\left(-\frac{x \alpha_{3,k}^{(j)}}{\alpha_{2,k}^{(j)} \tilde{\delta}_{SR,j}^2}\right) \right.$$
$$\times \sum_{m=0}^{\tau_1-1} \frac{x^m}{(\alpha_{2,k}^{(j)} \tilde{\delta}_{SR,j}^2)^m m!} \sum_{c=0}^{m} \binom{m}{c} (\alpha_{3,k}^{(j)})^{m-c}$$
$$\left. \left. \times \mathcal{J}\left(\tau_2+c; \frac{1}{\tilde{\delta}_{SP,j,k}^2} + \frac{x}{\alpha_{2,k}^{(j)} \tilde{\delta}_{SR,j}^2}\right) \right\} \right], \quad (13)$$

$$f_{\gamma_{SR}^{(j)}[n,k]}(x)$$
$$= \frac{1}{\Gamma(\tau_1)\Gamma(\tau_2)(\tilde{\delta}_{SR,j}^2)^{\tau_1}} \left[ \frac{1}{(\alpha_{1,k}^{(j)})^{\tau_1}} \gamma\left(\tau_2, \frac{P_A}{\tilde{\delta}_{SP,j,k}^2 P_M^{(j)}}\right) \right.$$
$$\times x^{\tau_1-1} \exp\left(-\frac{x}{\alpha_{1,k}^{(j)} \tilde{\delta}_{SR,j}^2}\right) + \frac{1}{(\alpha_{2,k}^{(j)})^{\tau_1}(\tilde{\delta}_{SP,j,k}^2)^{\tau_2}}$$
$$\times x^{\tau_1-1} \exp\left(-\frac{x \alpha_{3,k}^{(j)}}{\alpha_{2,k}^{(j)} \tilde{\delta}_{SR,j}^2}\right) \sum_{c=0}^{\tau_1} \binom{\tau_1}{c} (\alpha_{3,k}^{(j)})^{\tau_1-c}$$
$$\left. \times \mathcal{J}\left(\tau_2+c; \frac{x}{\alpha_{2,k}^{(j)} \tilde{\delta}_{SR,j}^2} + \frac{1}{\tilde{\delta}_{SP,j,k}^2}\right) \right], \quad (14)$$

where $\mathcal{J}(a;b) = \frac{\Gamma(a)}{b^a} \exp\left(-\frac{bP_A}{P_M^{(j)}}\right) \sum_{l=0}^{a-1} \frac{1}{l!} \left(\frac{bP_A}{P_M^{(j)}}\right)^l$, $\gamma(\cdot,\cdot)$ is the lower incomplete Gamma function[3] [39, Eq. (8.350.1)], and the various terms are defined as, $\tau_1 = N_S N_R$, $\tau_2 = N_S N_P$, $\tilde{\delta}_{SR,j}^2 = \delta_{SR,j}^2 + \sigma_{\epsilon_{SR},j}^2$, and $\tilde{\delta}_{SP,j,k}^2 = \rho_{SP,j}^{2(k-1)} \delta_{SP,j}^2 + \rho_{SP,j}^{2(k-1)} \sigma_{\epsilon_{SP},j}^2$.

*Proof:* Given in Appendix A. ∎

Now, using (4) and (13), the outage probability at the destination eNodeB corresponding to the transmission of a frame of $N_b$ codeword matrices by each of the $J$ SU-TXs can be derived as follows.

Let $\gamma_{\text{th}}$ denote the threshold for SNR outage at the destination node and $\epsilon(\gamma_{\text{th}})$ denote the corresponding outage event. Therefore, the net outage probability $\Pr(\epsilon(\gamma_{\text{th}}))$ for the cognitive network considering the mobile nature of the SU-TXs and PU-RXs, and imperfect channel estimates at the SU-

---

[3]The lower incomplete Gamma function $\gamma(s,x)$ with normalizing factor $\Gamma(s)$, i.e., $\frac{\gamma(s,x)}{\Gamma(s)}$ can be evaluated in MATLAB using the command gammainc($x,s$).

TXs and relay can be expressed as

$$\Pr(\epsilon(\gamma_{\text{th}})) = \frac{1}{JN_bB} \sum_{j=1}^{J} \sum_{k=1}^{N_b} \sum_{n=1}^{B} \Pr\left(\gamma_{\min}^{(j)}[n,k] \leq \gamma_{\text{th}}\right)$$
$$= \frac{1}{JN_bB} \sum_{j=1}^{J} \sum_{k=1}^{N_b} \sum_{n=1}^{B} F_{\gamma_{\min}^{(j)}[n,k]}(\gamma_{\text{th}}),$$

where $\gamma_{\min}^{(j)}[n,k] = \min\left\{\gamma_{SR}^{(j)}[n,k], \gamma_{RD}^{(j)}[n,k]\right\}$ denotes the end-to-end SNR and $F_{\gamma_{\min}^{(j)}[n,k]}(\cdot)$ represents the CDF of $\gamma_{\min}^{(j)}[n,k]$ given below [23, Eq. (11)]

$$F_{\gamma_{\min}^{(j)}[n,k]}(\gamma_{\text{th}}) = 1 - \left\{1 - F_{\gamma_{SR}^{(j)}[n,k]}(\gamma_{\text{th}})\right\}$$
$$\times \left\{1 - F_{\gamma_{RD}^{(j)}[n,k]}(\gamma_{\text{th}})\right\}, \quad (15)$$

where $F_{\gamma_{SR}^{(j)}[n,k]}(\gamma_{\text{th}})$ and $F_{\gamma_{RD}^{(j)}[n,k]}(\gamma_{\text{th}})$ denote the CDFs of the SNRs of the $j$th SU TX-relay and relay-eNodeB links and are given in (13) and (4) respectively. Further, the asymptotic outage analysis is given below.

### A. Asymptotic Outage Floor Analysis

In the high electrical SNR regime of the FSO link, the Meijer's G-function in (4) reduces to zero [46, Eq. (07.34.06.0001.01)]. Therefore, the outage floor corresponding to a high value of $\mu_\theta$ and a fixed value of $P_A$ can be determined as

$$\lim_{\mu_\theta \to \infty} \Pr(\epsilon(\gamma_{\text{th}})) = \frac{1}{JN_bB} \sum_{j=1}^{J} \sum_{k=1}^{N_b} \sum_{n=1}^{B} F_{\gamma_{SR}^{(j)}[n,k]}(\gamma_{\text{th}}), \quad (16)$$

where the CDF of the $j$th SU TX-relay link SNR $\gamma_{SR}^{(j)}[n,k]$ is given in (13). On the other hand, the outage floor corresponding to high average SNR values of the cognitive SU TX-relay RF links (when $P_A \to \infty$) and a fixed value of $\mu_\theta$ can be obtained as

$$\lim_{P_A \to \infty} \Pr(\epsilon(\gamma_{\text{th}})) = \frac{1}{JN_bB} \sum_{j=1}^{J} \sum_{k=1}^{N_b} \sum_{n=1}^{B} \left[1 - \left\{1 - F_{\gamma_{SR}^{(j)}[n,k]}(\gamma_{\text{th}})\right\}\right.$$
$$\left. \times \left\{1 - F_{\gamma_{RD}^{(j)}[n,k]}(\gamma_{\text{th}})\right\}\right], \quad (17)$$

where the CDF of received SNR $\gamma_{RD}^{(j)}[n,k]$ is given in (4) and the CDF of the SNR $\gamma_{SR}^{(j)}[n,k]$ for the $j$th SU TX-relay link when $P_A \to \infty$ can be derived using (12) as

$$F_{\gamma_{SR}^{(j)}[n,k]}(\gamma_{\text{th}}) = \Pr(\gamma_{SR}^{(j)}[n,k] \leq \gamma_{\text{th}})$$
$$= \Pr\left(G_{SR,L}^{(j)} \leq \frac{\gamma_{\text{th}}}{\alpha_{1,k}^{(j)}}\right)$$
$$= \frac{1}{\Gamma(\tau_1)} \gamma\left(\tau_1, \frac{\gamma_{\text{th}}}{\alpha_{1,k}^{(j)} \tilde{\delta}_{SR,j}^2}\right). \quad (18)$$

## IV. PROBABILITY OF ERROR ANALYSIS

Similar to outage analysis, the average probability of error for the cognitive network considering the mobile nature of the SU-TXs and PU-RXs, and imperfect channel estimates at the SU-TXs and relay can be expressed as

$$\overline{P}_e = \frac{1}{JN_bB} \sum_{j=1}^{J} \sum_{k=1}^{N_b} \sum_{n=1}^{B} P_e^{(j)}[n,k], \quad (19)$$

where $P_e^{(j)}[n,k]$ denotes the average probability of error for the $n$th symbol of the $k$th coded block transmitted by the $j$th SU-TX and can be derived using the CDF of $\gamma_{\min}^{(j)}[n,k]$ in (15) as [47, Eq. (21)]

$$P_e^{(j)}[n,k] = -\int_0^{\infty} P_e^{'}(x) F_{\gamma_{\min}^{(j)}[n,k]}(x) dx, \quad (20)$$

where $P_e^{'}(x)$ is the first order derivative of the conditional error probability. The conditional error probability for a given SNR $\gamma$ considering both coherent and non-coherent binary modulation schemes is [47, Eq. (25)]

$$P_e(\gamma) = \frac{\Gamma(a, b\gamma)}{2\Gamma(a)}, \quad (21)$$

where $\Gamma(\cdot,\cdot)$ is the upper incomplete Gamma function [39, Eq. (8.350.2)] and the constants $(a,b)$ are defined as, $(a,b) = (0.5, 1)$ for binary phase shift keying (BPSK), $(a,b) = (0.5, 0.5)$ for binary frequency shift keying (BFSK), $(a,b) = (1, 1)$ for Differential BPSK (DBPSK) and $(a,b) = (1, 0.5)$ for non-coherent BFSK (NCBFSK) modulation schemes. Differentiating (21) with respect to $\gamma$ and substituting the resulting expression in (20), the average probability of error $P_e^{(j)}[n,k]$ can be written as

$$P_e^{(j)}[n,k] = \frac{b^a}{2\Gamma(a)} \int_0^{\infty} e^{-bx} x^{a-1} F_{\gamma_{\min}^{(j)}[n,k]}(x) dx. \quad (22)$$

Employing (15), the above expression can be further simplified as

$$P_e^{(j)}[n,k] = \frac{b^a}{2\Gamma(a)} \left[ \underbrace{\int_0^{\infty} e^{-bx} x^{a-1} F_{\gamma_{SR}^{(j)}[n,k]}(x) dx}_{\triangleq \mathcal{L}_1} \right.$$

$$+ \underbrace{\int_0^{\infty} e^{-bx} x^{a-1} F_{\gamma_{RD}^{(j)}[n,k]}(x) dx}_{\triangleq \mathcal{L}_2}$$

$$\left. - \underbrace{\int_0^{\infty} e^{-bx} x^{a-1} F_{\gamma_{SR}^{(j)}[n,k]}(x) F_{\gamma_{RD}^{(j)}[n,k]}(x) dx}_{\triangleq \mathcal{L}_3} \right], \quad (23)$$

where the integral $\mathcal{L}_2$ can be directly simplified using the integral identity [39, Eq. (7.813.1)] as

$$\mathcal{L}_2 \triangleq \Theta_1 \int_0^{\infty} e^{-bx} x^{a-1} G_{\theta+1, 3\theta+1}^{3\theta, 1} \left( \frac{\Theta_2}{\mu_\theta} x \middle| \begin{array}{c} 1, \Theta_3 \\ \Theta_4, 0 \end{array} \right) dx$$

$$= \Theta_1 b^{-a} G_{\theta+2, 3\theta+1}^{3\theta, 2} \left( \frac{\Theta_2}{\mu_\theta b} \middle| \begin{array}{c} -a+1, 1, \Theta_3 \\ \Theta_4, 0 \end{array} \right), \quad (24)$$

and the closed-form expressions for integrals $\mathcal{L}_1$ and $\mathcal{L}_3$, as derived in Appendix B, are given in (25) and (26) respectively.

The function $G_{-,-;-;-,-;-,-}^{-,-;-;-,-;-,-}(-)$ in (26) is the Generalized Meijer's G-function of two variables[4] [50], [51]. This function is an extension of the Meijer's G-function and has been widely utilized to represent the product of three Meijer's G-functions in a closed-form.

### A. Asymptotic Error Floor

In the high electrical SNR regime of the FSO link, one can neglect the terms $\mathcal{L}_2$ and $\mathcal{L}_3$ in (23) since all the Meijer's G-functions in (24) and (26) reduce to zero [46, Eq. (07.34.06.0001.01)] when $\mu_\theta$ approaches $\infty$. Therefore, the error floor corresponding to a high value of $\mu_\theta$ and a fixed value of $P_A$ can be determined by considering the dominant term $\mathcal{L}_1$ in (23) and substituting the resulting expression for $P_e^{(j)}[n,k]$ in (19) as

$$\lim_{\mu_\theta \to \infty} \overline{P}_e = \frac{b^a}{2JN_bB\Gamma(b)} \sum_{j=1}^{J} \sum_{k=1}^{N_b} \sum_{n=1}^{B} \mathcal{L}_1, \quad (27)$$

where the closed-form expression for $\mathcal{L}_1$ is given in (25). On the other hand, the error floor corresponding to high average SNR values of the cognitive SU TX-relay RF links (when $P_A \to \infty$) and a fixed value of $\mu_\theta$ can be obtained as

$$\lim_{P_A \to \infty} \overline{P}_e = \frac{b^a}{2JN_bB\Gamma(b)} \sum_{j=1}^{J} \sum_{k=1}^{N_b} \sum_{n=1}^{B} \left[ \Gamma(a) b^{-a} - \sum_{l=0}^{\tau_1-1} \frac{1}{l!} \right.$$

$$\times \frac{1}{(\alpha_{1,k}^{(j)} \tilde{\delta}_{SR,j}^2)^l} \kappa^{-a-l} \left\{ \Gamma(a+l) - \Theta_1 \right.$$

$$\left. \left. \times G_{\theta+2, 3\theta+1}^{3\theta, 2} \left( \frac{\Theta_2}{\mu_\theta} \kappa^{-1} \middle| \begin{array}{c} -a-l+1, 1, \Theta_3 \\ \Theta_4, 0 \end{array} \right) \right\} \right], \quad (28)$$

where $\kappa = \left( b + \frac{1}{\alpha_{1,k}^{(j)} \tilde{\delta}_{SR,j}^2} \right)$. The detailed derivation for (28) is given in Appendix C.

## V. SIMULATION RESULTS

For simulation purposes, the transmission of the rate $R_c = \frac{1}{2}$ code $\mathcal{G}_c^3$ [8] is considered by $J = 5$ SU-TXs with the other parameters set as follows. Number of antennas $N_S = 3$, $N_R = N_P = 2$, average gains $\tilde{\delta}_{SR,j}^2 = \tilde{\delta}_{SP,j}^2 = 1, \forall j$, SNR outage threshold $\gamma_{\text{th}} = 3$ dB, frame length $N_b = 50$, peak transmit power $P_M^{(j)} = 27$ dBW, $\forall j$, and noise plus interference power $\eta_0 = 1$, which is valid for large scale networks when the PU-TX is located far away from the secondary network [3], [4], [41], [52], [53]. The correlation parameter $\rho_i$ corresponding to the relative speed of $\nu_i$ can be obtained by using the standard Jake's model [9] as, $\rho_i = \mathcal{J}_0 \left( \frac{2\pi f_c \nu_i}{R_s c} \right)$, where $\mathcal{J}_0(\cdot)$ is the zeroth-order Bessel function of the first kind, the carrier frequency and symbol rate are set as $f_c = 5.9$ GHz and $R_s = 9.5$ kbps respectively. Further, similar to [23], it is assumed that the FSO link experiences atmospheric turbulence with $C_n^2 \in \{3 \times 10^{-14}, 1 \times 10^{-13}\}$ m$^{-2/3}$, link length $L \in \{1, 2\}$ km, $\theta = 1$, aperture radius $r = 0.1$ m, and laser wavelength $\lambda = 1.55 \times 10^{-6}$ m. Under these conditions, the various parameters for

$$\mathcal{L}_1 = \frac{b^{-a}}{\Gamma(\tau_1)\Gamma(\tau_2)} \gamma\left(\tau_2, \frac{P_A}{\tilde{\delta}^2_{SP,j,k} P_M^{(j)}}\right) G_{2,2}^{1,2}\left(\frac{1}{\alpha_{1,k}^{(j)} \tilde{\delta}^2_{SR,j} b} \left|\begin{array}{c} -a+1, 1 \\ \tau_1, 0 \end{array}\right.\right) + \frac{\Gamma(a) b^{-a}}{(\tilde{\delta}^2_{SP,j,k})^{\tau_2} \Gamma(\tau_2)} \mathcal{J}\left(\tau_2; \frac{1}{\tilde{\delta}^2_{SP,j,k}}\right) - \frac{1}{\Gamma(\tau_2)}$$
$$\times \sum_{m=0}^{\tau_1-1} \frac{1}{(\alpha_{2,k}^{(j)} \tilde{\delta}^2_{SR,j})^m m!} \sum_{c=0}^{m} \binom{m}{c} (\alpha_{3,k}^{(j)})^{m-c} \Gamma(\tau_2+c) \exp\left(-\frac{P_A}{\tilde{\delta}^2_{SP,j,k} P_M^{(j)}}\right) \sum_{l=0}^{\tau_2+c-1} \frac{1}{l!} \left(\frac{P_A}{P_M^{(j)}}\right)^l \frac{(\tilde{\delta}^2_{SP,j,k})^{c-l}}{\Gamma(\tau_2+c-l)}$$
$$\times \left(\frac{\alpha_{3,k}^{(j)}}{\alpha_{2,k}^{(j)} \tilde{\delta}^2_{SR,j}} + \frac{P_A}{P_M^{(j)} \alpha_{2,k}^{(j)} \tilde{\delta}^2_{SR,j}} + b\right)^{-a-m} G_{2,1}^{1,2}\left(\tilde{\delta}^2_{SP,j,k}\left(\alpha_{3,k}^{(j)} + b\alpha_{2,k}^{(j)} \tilde{\delta}^2_{SR,j} + \frac{P_A}{P_M^{(j)}}\right)^{-1} \left|\begin{array}{c} -a-m+1, -\tau_2-c+l+1 \\ 0 \end{array}\right.\right). \quad (25)$$

---
[4]The complete Mathematica and MATLAB implementation codes for the Generalized Meijer's G-function of two variables can be found in [48] and [49] respectively.

---

$$\mathcal{L}_3 = \frac{\Theta_1 (\alpha_{1,k}^{(j)} \tilde{\delta}^2_{SR,j})^a}{\Gamma(\tau_1)\Gamma(\tau_2)} \gamma\left(\tau_2, \frac{P_A}{\tilde{\delta}^2_{SP,j,k} P_M^{(j)}}\right) G_{2,1:0,1:\theta+1,3\theta+1}^{1,1:1,0:3\theta,1}\left(\begin{array}{c}-a-\tau_1+1, -a+1 \\ -a\end{array}\left|\begin{array}{c}- \\ 0\end{array}\right|\begin{array}{c}1, \Theta_3 \\ \Theta_4, 0\end{array}\right| b\alpha_{1,k}^{(j)} \tilde{\delta}^2_{SR,j}, \frac{\Theta_2 \alpha_{1,k}^{(j)} \tilde{\delta}^2_{SR,j}}{\mu_\theta}\right)$$
$$+ \frac{\Theta_1 b^{-a}}{(\tilde{\delta}^2_{SP,j,k})^{\tau_2} \Gamma(\tau_2)} \mathcal{J}\left(\tau_2; \frac{1}{\tilde{\delta}^2_{SP,j,k}}\right) G_{\theta+2,3\theta+1}^{3\theta,2}\left(\frac{\Theta_2}{\mu_\theta b}\left|\begin{array}{c}-a+1, 1, \Theta_3 \\ \Theta_4, 0\end{array}\right.\right) - \frac{\Theta_1}{\Gamma(\tau_2)} \sum_{m=0}^{\tau_1-1} \frac{1}{(\alpha_{2,k}^{(j)} \tilde{\delta}^2_{SR,j})^m m!}$$
$$\times \sum_{c=0}^{m} \binom{m}{c} (\alpha_{3,k}^{(j)})^{m-c} \Gamma(\tau_2+c) \exp\left(-\frac{P_A}{\tilde{\delta}^2_{SP,j,k} P_M^{(j)}}\right) \sum_{l=0}^{\tau_2+c-1} \frac{1}{l!} \left(\frac{P_A}{P_M^{(j)}}\right)^l \frac{(\tilde{\delta}^2_{SP,j,k})^{c-l}}{\Gamma(\tau_2+c-l)} \left(\frac{\alpha_{2,k}^{(j)} \tilde{\delta}^2_{SR,j}}{\tilde{\delta}^2_{SP,j,k}}\right)^{a+m}$$
$$\times G_{1,1:0,1:\theta+1,3\theta+1}^{1,1:1,0:3\theta,1}\left(\begin{array}{c}-a-m+1 \\ -a-m-l+\tau_2+c\end{array}\left|\begin{array}{c}- \\ 0\end{array}\right|\begin{array}{c}1, \Theta_3 \\ \Theta_4, 0\end{array}\right| \frac{1}{\tilde{\delta}^2_{SP,j,k}}\left(b\alpha_{2,k}^{(j)} \tilde{\delta}^2_{SR,j} + \alpha_{3,k}^{(j)} + \frac{P_A}{P_M^{(j)}}\right), \frac{\Theta_2 \alpha_{2,k}^{(j)} \tilde{\delta}^2_{SR,j}}{\mu_\theta \tilde{\delta}^2_{SP,j,k}}\right). \quad (26)$$

---

the FSO link with moderate and strong atmospheric turbulence are set as, ($\alpha$=5.4181, $\beta$=3.7916, $H_l$=0.9033, $\xi$=1.6758) and ($\alpha$=5.0711, $\beta$=1.1547, $H_l$=0.8159, $\xi$=1.6885) respectively.

Fig. 2(a) shows the outage probability versus the received power limit $P_A$ at the PU-RXs for different speeds of the SU-TXs and PU-RXs with $\sigma^2_{e_{SR,j}} = 1, \forall j$ and perfect channel estimates. It can be observed that the system performance degrades for the scenario when only the SU-TXs are mobile as compared to the one in which each of the SU-TX and PU-RX nodes are static. Moreover, the probability of outage increases with the speed of the SU-TXs since higher speeds result in lower values of the correlation parameter $\rho_{SR,j}$ leading to a significant reduction in the effective SNRs at the relay. On the other hand, interestingly, the system performance can be seen to be better for the scenario with only the PU-RXs mobile in comparison to the scenarios with either the SU-TXs mobile or each SU-TX and PU-RX static. Also, the outage probabilities in the low and moderate SNR regimes further decrease with increasing speed of the PU-RXs. This interesting result is due to the fact that high speed of the PU-RX results in a progressive reduction in the cross channel gain of the time-selective SU TX-PU RX link as can be seen from Section II. Hence, from the comprehensive power constraint in (5) it follows that the transmit power of the underlay SU can be increased without compromising the performance of the primary user. One can also observe that OSTBC based transmissions over the MIMO SU TX-relay links with DF cooperation outperforms the SISO system with variable gain relaying proposed in [2] in moderate and high SNR regimes. However, the system performance with either the SU-TXs mobile or each SU-TX and PU-TX static is marginally worse in the low SNR regime due to power normalization by the factor $R_C N_S$ where $N_S$=3 and $R_C$=1/2. Moreover, for the scenario when only the PU-RXs are mobile, significant performance improvement can also be seen in the low SNR regime. Further, a key utility of the above results in the context of practical system design is to determine a suitable interference margin in the link budget for the primary-user to meet a desired level of outage probability in a fading channel. Since the transmit power of the secondary users is limited by the interference threshold of the primary users, it cannot be increased arbitrarily to include a fade margin and is tied to the interference threshold $P_A$ of the primary users. Hence, the required interference margins in both the moderate and strong turbulence regimes and the corresponding results are described in detail in the technical report in [35].

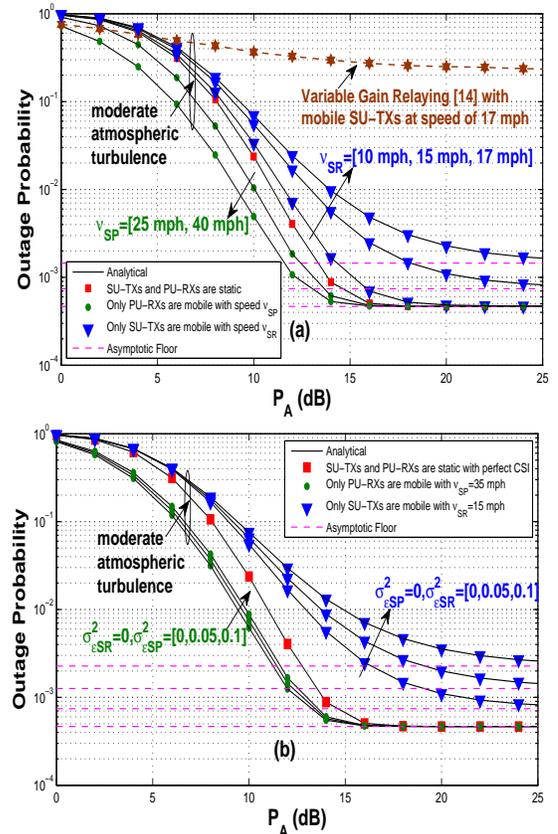

Fig. 2. Outage performance of a cognitive DF MIMO-RF/FSO DF cooperative system with average relay-eNodeB SNR $\bar{\gamma}_{RD} = 20$ dB.

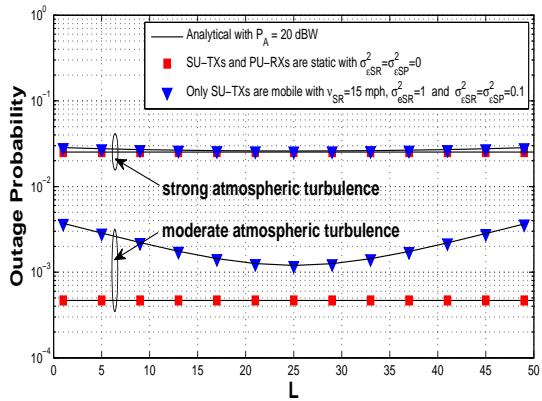

Fig. 3. Outage performance of a cognitive mixed MIMO-RF/FSO DF cooperative system with average relay-eNodeB SNR $\bar{\gamma}_{RD} = 20$ dB.

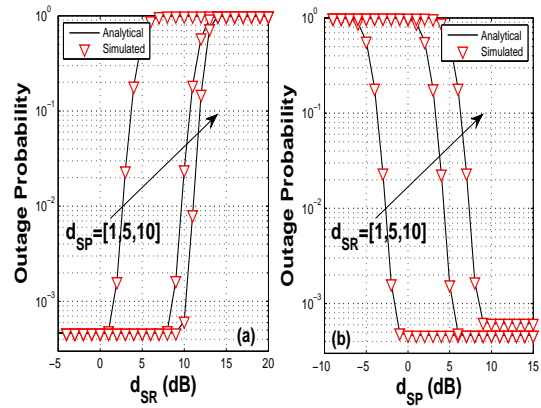

Fig. 4. Impact of distance on outage performance with moderate atmospheric turbulence, $P_A$=15 dB, $L$=1, $N_S$=$N_R$=3, $N_P$=2, and $\bar{\gamma}_{RD}$=20 dB.

Fig. 2(b) demonstrates the impact of the imperfect CSI obtained at the beginning of each frame i.e., $L = 1$, where $\sigma^2_{\epsilon_{SR,j}} = \sigma^2_{\epsilon_{SR}}, \sigma^2_{\epsilon_{SP,j}} = \sigma^2_{\epsilon_{SP}}, \forall j$. It can be observed that the outage probability for the scenario with only the SU-TXs mobile increases with increasing variance of channel estimation error at the relay. Further, in contrast to the previous observation wherein high mobile speed of the PU-RXs results in a performance improvement, imperfect channel estimates of the time-selective SU TX-PU RX link at the SU-TX degrade the outage performance. Moreover, the probability of outage further increases in the low and moderate SNR regimes as the variance of channel estimation error i.e., $\sigma^2_{\epsilon_{SP}}$ increases. The variance of channel estimation error i.e., $\sigma^2_{\epsilon_{SR}}$ can be seen to have a significant impact on the outage probability. At outage probability $2 \times 10^{-3}$, the interference threshold at the primary user has to be raised by approximately 3 dB as $\sigma^2_{\epsilon_{SR}}$ increases from 0 to 0.05. Therefore, it is essential to design efficient schemes for channel estimation as well as pilot placement, which reduce the deterioration of the system outage probability for the same pilot overhead.

Fig. 3 illustrates the effect of preamble versus midamble for channel estimation at the relay where the midamble i.e., $L = 25$ can be seen to result in a significantly lower outage probability in comparison to the preamble i.e., $L = 1$ considered in [28]–[32], [44], [54]. This is owing to the fact that in contrast to midamble, the error in preamble progressively increases from $k = 1$ to $k = N_b$ due to the time selective nature of the link. Thus a midamble, which is a readily employable strategy, yields a substantial reduction in the outage probability in comparison to the preamble for the same pilot overhead. One can also observe that in comparison to moderate atmospheric turbulence, strong atmospheric turbulence in FSO link significantly degrades the end-to-end system performance.

Fig. 4 demonstrates the impact of distance on outage performance, where the presence of single SU-TX and PU-

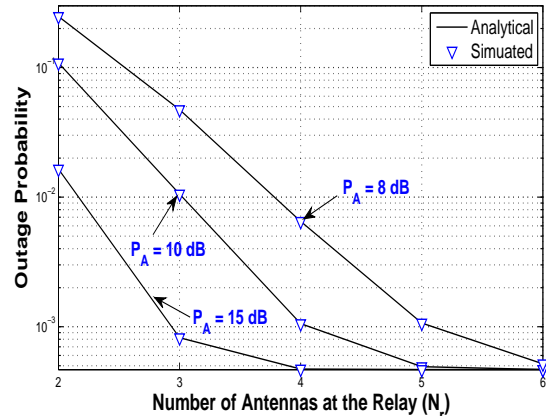

Fig. 5. Impact of number of antennas $(N_R)$ on outage performance with moderate atmospheric turbulence and average relay-eNodeB SNR $\bar{\gamma}_{RD} = 20$ dB.

RX nodes is considered with $\delta^2_{SR} = d^{-\alpha}_{SR}, \delta^2_{SP} = d^{-\alpha}_{SP}$, $\alpha = 2.5$ is the path loss exponent and $d_{SR}, d_{SP}$ denote the distances between the SU-TX and relay, and the SU-TX and PU-RX, respectively. Firstly, one can observe that the end-to-end system performance severely degrades as the distance $d_{SR}$ between the SU-TX and relay increases. It is worth noting that the system experiences outage with probability $1.6 \times 10^{-3}$ for $d_{SP} = 1$ and $d_{SR} = 2$. Interestingly, as the distance $d_{SP}$ between the SU-TX and PU-RX increases from 1 to 5, the system experiences the same outage for a higher value of $d_{SR} \approx 10.5$. This is owing to the fact that as $d_{SP}$ increases, it decreases the cross channel gain, which enables an increase in the SU-TX transmit power in the underlay mode that supports signal reception at a large distance $d_{SR}$ for the same outage. Further, it can also be seen in Fig. 4(b) that the outage performance significantly improves as $d_{SP}$ increases and experiences a floor for large values of $d_{SP}$ since the end-to-end performance is limited by the FSO link.

Fig. 5 demonstrates the impact of number of antennas $(N_R)$ at the relay on the end-to-end outage performance of the system for various values of the interference threshold $P_A$. For this simulation, the SU-TXs and PU-RXs are assumed to be mobile with speed 17 mph, $\sigma^2_{eSR,j} = 1, \forall j$, and imperfect channel estimates are considered at the relay, SU-TXs with $\sigma^2_{\epsilon_{SR,j}} = \sigma^2_{\epsilon_{SP,j}} = 0.1, \forall j$. It can be observed in Fig. 5 that the system performance can be significantly enhanced by increasing the number of antennas at the relay. For example, the outage probability with $P_A = 10$ dB reduces from $0.1$ to $0.001$ as $N_R$ increases from $2$ to $4$. However, the system experiences an outage floor as $N_R$ increases further. This is owing to the fact that the end-to-end system performance is dominated by the FSO link.

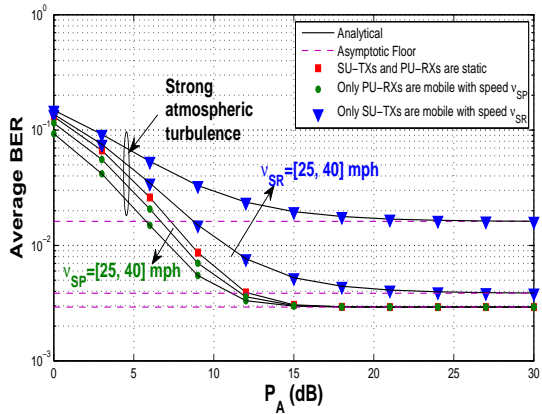

Fig. 6. Error performance of a cognitive DF MIMO-RF/FSO DF cooperative system with average relay-eNodeB SNR $\bar{\gamma}_{RD} = 20$ dB.

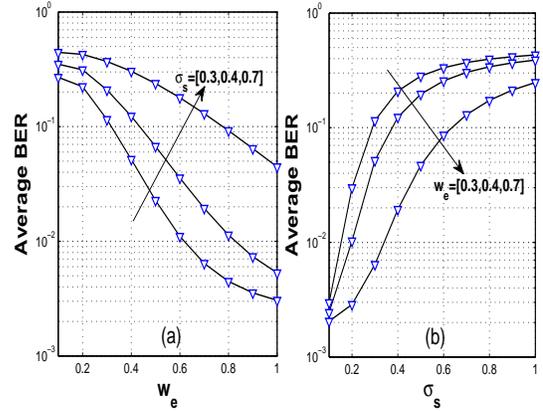

Fig. 7. Error performance of a cognitive DF MIMO-RF/FSO DF cooperative system with $\bar{\gamma}_{RD} = P_A = 20$ dB, strong turbulence ($\alpha = 5.0711, \beta = 1.1547, H_l = 0.8159$), and (a) fixed standard deviation of the pointing error displacement $\sigma_s$ (b) fixed beam-width radius $w_e$.

The error rate performance of the cognitive RF/FSO transmission with moderate atmospheric turbulence is demonstrated in Fig. 6 under various node mobility conditions, where imperfect channel estimates are assumed to be available using preamble with $\sigma^2_{\epsilon_{SR},j} = \sigma^2_{\epsilon_{SP},j} = 0.05, \forall j$. Firstly, it can be seen that the analytical values obtained using (19) exactly match with the simulated ones, thus validating the analytical framework derived. Second, similar to outage performance, improvement in the end-to-end probability of error can also be observed for the scenario when only the PU-RXs are mobile. Further, one can also observe that the cognitive system performance experiences the error floor derived in (28) at high values of $P_A$. For the scenarios when either each of the SU-TX and PU-RX nodes are static or only PU-RXs are mobile, the system experiences an asymptotic floor at high $P_A$ due to the weak FSO link, as can be seen in Fig. 6. However, for the scenario when only the SU-TXs are mobile with 25 or 45 mph, the end-to-end system performance is dominated by the cognitive RF link and experiences the floor due to SU-TX mobility.

The impact of FSO link pointing errors on the end-to-end error rate performance is further analyzed in Fig. 7, where the SU-TXs are assumed to be mobile with a speed of 40 mph and only imperfect channel estimates are available at the SU-TXs and relay with $\sigma^2_{\epsilon_{SR},j} = \sigma^2_{\epsilon_{SP},j} = 0.05, \forall j$ at the beginning of each frame. It can be clearly seen that end-to-end performance significantly degrades with the increase in standard deviation $\sigma_s$ of the pointing error displacement. On the other hand, the performance can be significantly enhanced by increasing the beam-width radius $w_e$.

Finally, Fig. 8 demonstrates the impact of cognitive transmission on the outage performance of the primary $2 \times 2$ MIMO network wherein the Alamouti code is used for transmission with fixed transmit power $P_U = 20$ dB. It can be clearly seen that the outage performance of the primary network significantly degrades as the peak interference power $P_A$ increases. This is due to the fact that a high value of $P_A$ results in high interference at the primary receiver. Fig. 8 also compares the performance with that of our existing scheme proposed in [23], which considers only the fixed interference power constraint for the cognitive transmission, while neglecting the proportional interference power constraint. The proposed scheme leads to an improvement in the outage performance of the primary communication system in comparison to the existing scheme as can be clearly seen in Fig. 8.

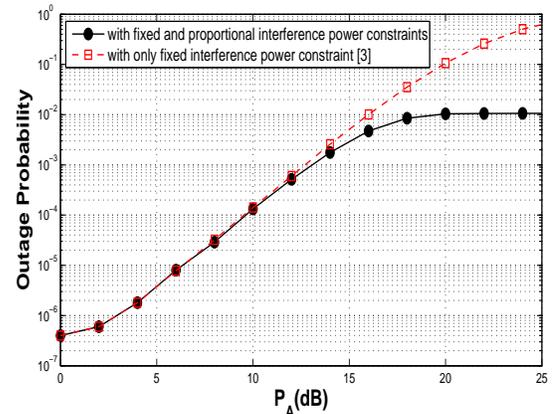

Fig. 8. Outage performance of the primary network in the presence of cognitive transmission with static nodes, perfect channel estimates, $P_M = 10$ dBW, and fixed primary transmit power $P_U = 20$ dBW.

## VI. CONCLUSION

This work analyzed the performance of an underlay cognitive mixed MIMO-RF/FSO DF cooperative relay network considering time-selective fading MIMO RF links as well as imperfect channel estimates at the relay and each of the SU-TXs. It was shown that the mobile nature of the SU-TXs significantly degrades the system performance. However, time-selective fading in the SU TX-PU RX RF links, arising due to the mobile nature of the PU-RXs, results in performance improvement. This interesting result is due to the fact that high speed of the PU-RX results in a progressive reduction in the cross channel gain of the time-selective SU TX-PU RX link which in turns increases the transmit power of the underlay SU. Moreover, it is also shown that a midamble yields a substantial performance improvement in comparison to a preamble. This is owing to the fact that in contrast to a midamble, the CSI error from a preamble progressively increases over the block duration due to the time selective nature of the link. On the other hand, an improved primary network outage performance can be seen in comparison to the existing scheme that neglects the proportional interference power constraint. Finally, future studies can focus on the opportunistic scheduling of multiple SU-TXs as well as the optimal metric for selection of the best SU-TX for transmission with node mobility and imperfect CSI.

## APPENDIX A
## CDF AND PDF OF $j$TH SU TX-RELAY LINK SNR $\gamma_{SR}^{(j)}[n,k]$

The CDF of $\gamma_{SR}^{(j)}[n,k]$ given in (12) can be obtained as

$$F_{\gamma_{SR}^{(j)}[n,k]}(x)$$
$$= \int_0^{\frac{P_A}{P_M^{(j)}}} \left( \int_0^{\frac{x}{\alpha_{1,k}^{(j)}}} f_{G_{SR,L}^{(j)}}(y) dy \right) f_{G_{SP,k}^{(j)}}(z) dz$$
$$+ \int_{\frac{P_A}{P_M^{(j)}}}^{\infty} \left( \int_0^{\frac{x(z+\alpha_{3,k}^{(j)})}{\alpha_{2,k}^{(j)}}} f_{G_{SR,L}^{(j)}}(y) dy \right) f_{G_{SP,k}^{(j)}}(z) dz, \quad (29)$$
$$= F_{G_{SR,L}^{(j)}}\left(\frac{x}{\alpha_{1,k}^{(j)}}\right) F_{G_{SP,k}^{(j)}}\left(\frac{P_A}{P_M^{(j)}}\right)$$
$$+ \underbrace{\int_{\frac{P_A}{P_M^{(j)}}}^{\infty} F_{G_{SR,L}^{(j)}}\left(\frac{x(z+\alpha_{3,k}^{(j)})}{\alpha_{2,k}^{(j)}}\right) f_{G_{SP,k}^{(j)}}(z) dz}_{\triangleq \mathcal{I}_1}, \quad (30)$$

where $F_{G_{SR,L}^{(j)}}(\cdot)$ denotes the CDF of $G_{SR,L}^{(j)} = \|\widehat{\mathbf{H}}_{SR}^{(j)}[L]\|_F^2$ and is given as

$$F_{G_{SR,L}^{(j)}}(x) = \frac{1}{\Gamma(\tau_1)} \gamma\left(\tau_1, \frac{x}{\tilde{\delta}_{SR,j}^2}\right), \quad (31)$$

where $\tau_1 = N_S N_R$, $\tilde{\delta}_{SR,j}^2 = \delta_{SR,j}^2 + \sigma_{\epsilon_{SR,j}}^2$, and $\gamma(\cdot,\cdot)$ is the lower incomplete Gamma function [39, Eq. (8.350.1)]. The quantities $F_{G_{SP,k}^{(j)}}(\cdot)$ and $f_{G_{SP,k}^{(j)}}(\cdot)$ in (30) denote the CDF and PDF of $G_{SP,k}^{(j)} = \|\widehat{\mathbf{H}}_{SP}^{(j)}[k]\|_F^2$ respectively and are given as

$$F_{G_{SP,k}^{(j)}}(x) = \frac{1}{\Gamma(\tau_2)} \gamma\left(\tau_2, \frac{x}{\tilde{\delta}_{SP,j,k}^2}\right), \quad (32)$$

$$f_{G_{SP,k}^{(j)}}(x) = \frac{1}{(\tilde{\delta}_{SP,j,k}^2)^{\tau_2} \Gamma(\tau_2)} x^{\tau_2-1} \exp\left(-\frac{x}{\tilde{\delta}_{SP,j,k}^2}\right), \quad (33)$$

where $\tau_2 = N_S N_P$ and $\tilde{\delta}_{SP,j,k}^2 = \rho_{SP,j}^{2(k-1)} \delta_{SP,j}^2 + \rho_{SP,j}^{2(k-1)} \sigma_{\epsilon_{SP,j}}^2$. Employing the expressions (31), (32), the integral $\mathcal{I}_1$ in (30) can be written as

$$\mathcal{I}_1 = \frac{1}{(\tilde{\delta}_{SP,j,k}^2)^{\tau_2} \Gamma(\tau_2) \Gamma(\tau_1)} \int_{\frac{P_A}{P_M^{(j)}}}^{\infty} z^{\tau_2-1} \exp\left(-\frac{z}{\tilde{\delta}_{SP,j,k}^2}\right)$$
$$\times \gamma\left(\tau_1, \frac{x(z+\alpha_{3,k}^{(j)})}{\alpha_{2,k}^{(j)} \tilde{\delta}_{SR,j}^2}\right) dz. \quad (34)$$

Now, using the identity $\gamma(s,x) = (s-1)! - (s-1)! e^{-x} \sum_{m=0}^{s-1} \frac{x^m}{m!}$ from [39, Eq. (8.352.1)] along with $(s-1)! = \Gamma(s)$, the above expression can be rewritten as

$$\mathcal{I}_1 = \frac{1}{(\tilde{\delta}_{SP,j,k}^2)^{\tau_2} \Gamma(\tau_2)} \left[ \int_{\frac{P_A}{P_M^{(j)}}}^{\infty} z^{\tau_2-1} \exp\left(-\frac{z}{\tilde{\delta}_{SP,j,k}^2}\right) dz \right.$$
$$- \exp\left(-\frac{x \alpha_{3,k}^{(j)}}{\alpha_{2,k}^{(j)} \tilde{\delta}_{SR,j}^2}\right) \sum_{m=0}^{\tau_1-1} \frac{x^m}{m!} \int_{\frac{P_A}{P_M^{(j)}}}^{\infty} (z + \alpha_{3,k}^{(j)})^m$$
$$\left. \times z^{\tau_2-1} \exp\left(-\left(\frac{1}{\tilde{\delta}_{SP,j,k}^2} + \frac{x}{\alpha_{2,k}^{(j)} \tilde{\delta}_{SR,j}^2}\right) z\right) dz \right]. \quad (35)$$

Further, considering the binomial expansion for $(z + \alpha_{3,k}^{(j)})^m = \sum_{c=0}^{m} \binom{m}{c} z^c (\alpha_{3,k}^{(j)})^{m-c}$ and subsequently using the identity $\int_x^{\infty} t^{s-1} \exp(-\mu t) dt = \frac{\Gamma(s, \mu x)}{\mu^s}$ [39, Eq. (3.351.2)], the integrals above can be simplified to yield the closed-form expression for $\mathcal{I}_1$ as

$$\mathcal{I}_1 = \frac{1}{(\tilde{\delta}_{SP,j,k}^2)^{\tau_2} \Gamma(\tau_2)} \left[ \mathcal{J}\left(\tau_2; \frac{1}{\tilde{\delta}_{SP,j,k}^2}\right) \right.$$
$$- \exp\left(-\frac{x \alpha_{3,k}^{(j)}}{\alpha_{2,k}^{(j)} \tilde{\delta}_{SR,j}^2}\right) \sum_{m=0}^{\tau_1-1} \frac{x^m}{m!} \sum_{c=0}^{m} \binom{m}{c} (\alpha_{3,k}^{(j)})^{m-c}$$
$$\left. \times \mathcal{J}\left(\tau_2 + c; \frac{1}{\tilde{\delta}_{SP,j,k}^2} + \frac{x}{\alpha_{2,k}^{(j)} \tilde{\delta}_{SR,j}^2}\right) \right], \quad (36)$$

where the function $\mathcal{J}(a; b)$ is defined as, $\mathcal{J}(a; b) = \int_{\frac{P_A}{P_M^{(j)}}}^{\infty} z^{a-1} \exp(-bz) dz = \frac{1}{b^a} \Gamma\left(a, \frac{bP_A}{P_M^{(j)}}\right) = \frac{\Gamma(a)}{b^a} \exp\left(-\frac{bP_A}{P_M^{(j)}}\right) \times \sum_{l=0}^{a-1} \frac{1}{l!} \left(\frac{bP_A}{P_M^{(j)}}\right)^l$ [39, Eq. (3.352.2)], where $\Gamma(\cdot,\cdot)$ is the upper incomplete Gamma function [39, Eq. (8.350.2)]. Finally, substituting the above expression along with (31), (32) for $F_{G_{SP,k}^{(j)}}(x)$ and $F_{G_{SR,L}^{(j)}}(x)$ in (30), the closed form expression for the CDF of $j$th SU TX-relay link SNR $\gamma_{SR}^{(j)}[n,k]$ can be derived as (13).

On the other hand, the PDF of $\gamma_{SR}^{(j)}[n,k]$ can be obtained by differentiating (29) with respect to $x$ as

$$f_{\gamma_{SR}^{(j)}[n,k]}(x)$$
$$= \int_0^{\frac{P_A}{P_M^{(j)}}} \frac{1}{\alpha_{1,k}^{(j)}} f_{G_{SR,L}^{(j)}}\left(\frac{x}{\alpha_{1,k}^{(j)}}\right) f_{G_{SP,k}^{(j)}}(z) dz$$
$$+ \int_{\frac{P_A}{P_M^{(j)}}}^{\infty} \frac{(z + \alpha_{3,k}^{(j)})}{\alpha_{2,k}^{(j)}} f_{G_{SR,L}^{(j)}}\left(\frac{x(z+\alpha_{3,k}^{(j)})}{\alpha_{2,k}^{(j)}}\right) f_{G_{SP,k}^{(j)}}(z) dz,$$

$$= \frac{1}{\alpha_{1,k}^{(j)}} f_{G_{SR,L}^{(j)}} \left(\frac{x}{\alpha_{1,k}^{(j)}}\right) F_{G_{SP,k}^{(j)}} \left(\frac{P_A}{P_M^{(j)}}\right)$$
$$+ \underbrace{\int_{\frac{P_A}{P_M^{(j)}}}^{\infty} \frac{(z+\alpha_{3,k}^{(j)})}{\alpha_{2,k}^{(j)}} f_{G_{SR,L}^{(j)}} \left(\frac{x(z+\alpha_{3,k}^{(j)})}{\alpha_{2,k}^{(j)}}\right) f_{G_{SP,k}^{(j)}}(z) dz}_{\triangleq \mathcal{I}_2}, \quad (37)$$

where $f_{G_{SR,L}^{(j)}}(\cdot)$ denotes the PDF of $G_{SR,L}^{(j)} = \|\widehat{\mathbf{H}}_{SR}^{(j)}[L]\|_F^2$ and is given as

$$f_{G_{SR,L}^{(j)}}(x) = \frac{1}{(\tilde{\delta}_{SR,j}^2)^{\tau_1} \Gamma(\tau_1)} x^{\tau_1-1} \exp\left(-\frac{x}{\tilde{\delta}_{SR,j}^2}\right). \quad (38)$$

Employing the above expression along with (33), the integral $\mathcal{I}_2$ in (37) can be written as

$$\mathcal{I}_2 = \frac{x^{\tau_1-1} \exp(-x\alpha_{3,k}^{(j)}/(\alpha_{2,k}^{(j)}\tilde{\delta}_{SR,j}^2))}{(\alpha_{2,k}^{(j)}\tilde{\delta}_{SR,j}^2)^{\tau_1} (\tilde{\delta}_{SP,j,k}^2)^{\tau_2} \Gamma(\tau_2) \Gamma(\tau_1)} \int_{\frac{P_A}{P_M^{(j)}}}^{\infty} (z+\alpha_{3,k}^{(j)})^{\tau_1}$$
$$\times z^{\tau_2-1} \exp\left(-\left(\frac{x}{\alpha_{2,k}^{(j)}\tilde{\delta}_{SR,j}^2} + \frac{1}{\tilde{\delta}_{SP,j,k}^2}\right)z\right) dz. \quad (39)$$

Now, considering the binomial expansion for $(z+\alpha_{3,k}^{(j)})^{\tau_1} = \sum_{c=0}^{\tau_1} \binom{\tau_1}{c} z^c (\alpha_{3,k}^{(j)})^{\tau_1-c}$ and subsequently using the identity $\int_x^{\infty} t^{s-1} \exp(-\mu t) dt = \frac{\Gamma(s,\mu x)}{\mu^s}$ [39, Eq. (3.351.2)], the integrals above can be simplified to yield the closed-form expression for $\mathcal{I}_2$ as

$$\mathcal{I}_2 = \frac{x^{\tau_1-1} \exp(-x\alpha_{3,k}^{(j)}/(\alpha_{2,k}^{(j)}\tilde{\delta}_{SR,j}^2))}{(\alpha_{2,k}^{(j)}\tilde{\delta}_{SR,j}^2)^{\tau_1} (\tilde{\delta}_{SP,j,k}^2)^{\tau_2} \Gamma(\tau_2) \Gamma(\tau_1)} \sum_{c=0}^{\tau_1} \binom{\tau_1}{c} (\alpha_{3,k}^{(j)})^{\tau_1-c}$$
$$\times \mathcal{J}\left(\tau_2+c; \frac{x}{\alpha_{2,k}^{(j)}\tilde{\delta}_{SR,j}^2} + \frac{1}{\tilde{\delta}_{SP,j,k}^2}\right). \quad (40)$$

Finally, substituting the above expression along with (32), (38) for $F_{G_{SP,k}^{(j)}}(x)$ and $f_{G_{SR,L}^{(j)}}(x)$ in (37), the closed form expression for the PDF of $j$th SU TX-relay link SNR $\gamma_{SR}^{(j)}[n,k]$ can be derived as (14).

## APPENDIX B
### SIMPLIFICATION FOR INTEGRALS $\mathcal{L}_1$ AND $\mathcal{L}_3$

Substituting (13), the integral $\mathcal{L}_1$ in (23) can be written as

$$\mathcal{L}_1 = \frac{1}{\Gamma(\tau_1)\Gamma(\tau_2)} \gamma\left(\tau_2, \frac{P_A}{P_M^{(j)}\tilde{\delta}_{SP,j,k}^2}\right) \int_0^{\infty} x^{a-1} \exp(-bx)$$
$$\times \gamma\left(\tau_1, \frac{x}{\alpha_{1,k}^{(j)}\tilde{\delta}_{SR,j}^2}\right) dx + \frac{(\tilde{\delta}_{SP,j,k}^2)^{-\tau_2}}{\Gamma(\tau_2)} \mathcal{J}\left(\tau_2; \frac{1}{\tilde{\delta}_{SP,j,k}^2}\right)$$
$$\times \int_0^{\infty} x^{a-1} \exp(-bx) dx - \frac{1}{\Gamma(\tau_2)(\tilde{\delta}_{SP,j,k}^2)^{\tau_2}}$$
$$\times \sum_{m=0}^{\tau_1-1} \frac{1}{(\alpha_{2,k}^{(j)}\tilde{\delta}_{SR,j}^2)^m m!} \sum_{c=0}^{m} \binom{m}{c} (\alpha_{3,k}^{(j)})^{m-c} \Gamma(\tau_2+c)$$
$$\times \exp\left(-\frac{P_A}{\tilde{\delta}_{SP,j,k}^2 P_M^{(j)}}\right) \sum_{l=0}^{\tau_2+c-1} \frac{1}{l!} \left(\frac{P_A}{P_M^{(j)}}\right)^l (\tilde{\delta}_{SP,j,k}^2)^{\tau_2+c-l}$$
$$\times \int_0^{\infty} x^{a+m-1} \left(1 + \frac{\tilde{\delta}_{SP,j,k}^2 x}{\alpha_{2,k}^{(j)}\tilde{\delta}_{SR,j}^2}\right)^{l-c-\tau_2}$$
$$\times \exp\left(-\left(\frac{\alpha_{3,k}^{(j)}}{\alpha_{2,k}^{(j)}\tilde{\delta}_{SR,j}^2} + \frac{P_A}{P_M^{(j)}\alpha_{2,k}^{(j)}\tilde{\delta}_{SR,j}^2} + b\right)x\right) dx. \quad (41)$$

Further, using the identities $\gamma(\nu,x) = G_{1,2}^{1,1}\left(x \,\middle|\, \begin{array}{c} 1 \\ \nu, 0 \end{array}\right)$ [55, Eq. (8.4.16.1)] and $(1+x)^{-\beta} = \frac{1}{\Gamma(\beta)} G_{1,1}^{1,1}\left(x \,\middle|\, \begin{array}{c} 1-\beta \\ 0 \end{array}\right)$ [55, Eq. (8.4.2.5)] along with [39, Eq. (7.813.1)]

$$\int_0^{\infty} x^{-\rho} \exp(-\beta x) G_{p,q}^{m,n}\left(\alpha x \,\middle|\, \begin{array}{c} a_1, a_2, \cdots, a_p \\ b_1, b_2, \cdots, b_q \end{array}\right) dx$$
$$= \beta^{\rho-1} G_{p+1,q}^{m,n+1}\left(\frac{\alpha}{\beta} \,\middle|\, \begin{array}{c} \rho, a_1, a_2, \cdots, a_p \\ b_1, b_2, \cdots, b_q \end{array}\right), \quad (42)$$

the equation in (41) can be simplified to yield the final expression for $\mathcal{L}_1$ in (25).

On the other hand, the integral $\mathcal{L}_3$ in (23) can be simplified after substituting the expressions (4) and (13) for $F_{\gamma_{RD}^{(j)}[n,k]}(x)$ and $F_{\gamma_{SR}^{(j)}[n,k]}(x)$, respectively, as

$$\mathcal{L}_3 = \frac{\Theta_1}{\Gamma(\tau_1)\Gamma(\tau_2)} \gamma\left(\tau_2, \frac{P_A}{P_M^{(j)}\tilde{\delta}_{SP,j,k}^2}\right) \int_0^{\infty} x^{a-1} \exp(-bx)$$
$$\times \gamma\left(\tau_1, \frac{x}{\alpha_{1,k}^{(j)}\tilde{\delta}_{SR,j}^2}\right) G_{\theta+1,3\theta+1}^{3\theta,1}\left(\frac{\Theta_2}{\mu_\theta} x \,\middle|\, \begin{array}{c} 1, \Theta_3 \\ \Theta_4, 0 \end{array}\right) dx$$
$$+ \frac{\Theta_1}{(\tilde{\delta}_{SP,j,k}^2)^{\tau_2} \Gamma(\tau_2)} \mathcal{J}\left(\tau_2; \frac{1}{\tilde{\delta}_{SP,j,k}^2}\right) \int_0^{\infty} x^{a-1} \exp(-bx)$$
$$\times G_{\theta+1,3\theta+1}^{3\theta,1}\left(\frac{\Theta_2}{\mu_\theta} x \,\middle|\, \begin{array}{c} 1, \Theta_3 \\ \Theta_4, 0 \end{array}\right) dx - \frac{\Theta_1}{\Gamma(\tau_2)(\tilde{\delta}_{SP,j,k}^2)^{\tau_2}}$$
$$\times \sum_{m=0}^{\tau_1-1} \frac{1}{(\alpha_{2,k}^{(j)}\tilde{\delta}_{SR,j}^2)^m m!} \sum_{c=0}^{m} \binom{m}{c} (\alpha_{3,k}^{(j)})^{m-c} \Gamma(\tau_2+c)$$
$$\times \exp\left(-\frac{P_A}{\tilde{\delta}_{SP,j,k}^2 P_M^{(j)}}\right) \sum_{l=0}^{\tau_2+c-1} \frac{1}{l!} \left(\frac{P_A}{P_M^{(j)}}\right)^l (\tilde{\delta}_{SP,j,k}^2)^{\tau_2+c-l}$$
$$\times \int_0^{\infty} x^{a+m-1} \exp\left(-\left(\frac{\alpha_{3,k}^{(j)}}{\alpha_{2,k}^{(j)}\tilde{\delta}_{SR,j}^2} + \frac{P_A}{P_M^{(j)}\alpha_{2,k}^{(j)}\tilde{\delta}_{SR,j}^2} + b\right)x\right)$$
$$\times \left(1 + \frac{\tilde{\delta}_{SP,j,k}^2 x}{\alpha_{2,k}^{(j)}\tilde{\delta}_{SR,j}^2}\right)^{l-c-\tau_2} G_{\theta+1,3\theta+1}^{3\theta,1}\left(\frac{\Theta_2}{\mu_\theta} x \,\middle|\, \begin{array}{c} 1, \Theta_3 \\ \Theta_4, 0 \end{array}\right) dx.$$

Subsequently employing the identities for $\gamma(\nu,x)$ [55, Eq. (8.4.16.1)], $(1+x)^{-\beta}$ [55, Eq. (8.4.2.5)], and $\exp(-bx)$ [55, Eq. (8.4.3.1)] in terms of Meijer's G-function along with [46, Eq. (07.34.21.0081.01)], the above equation can be solved to yield the final expression for $\mathcal{L}_3$ in (26).

## APPENDIX C
### DERIVATION FOR ERROR FLOOR WHEN $P_A \to \infty$

Substituting the CDF $F_{\gamma_{SR}^{(j)}[n,k]}(x) = \frac{1}{\Gamma(\tau_1)} \gamma\left(\tau_1, \frac{x}{\alpha_{1,k}^{(j)}\tilde{\delta}_{SR,j}^2}\right)$ for the $j$th SU TX-relay link SNR from (18) along with the CDF $F_{\gamma_{RD}^{(j)}[n,k]}(x)$ of the FSO link SNR from (4) in (23), the

expression for $P_e^{(j)}[n,k]$ can be written as

$$P_e^{(j)}[n,k] = \frac{b^a}{2\Gamma(a)} \left[ \frac{1}{\Gamma(\tau_1)} \int_0^\infty e^{-bx} x^{a-1} \gamma\left(\tau_1, \frac{x}{\alpha_{1,k}^{(j)} \tilde{\delta}_{SR,j}^2}\right) dx \right.$$
$$+ \Theta_1 \int_0^\infty e^{-bx} x^{a-1} G_{\theta+1,3\theta+1}^{3\theta,1}\left(\frac{\Theta_2}{\mu_\theta} x \left| \begin{array}{c} 1, \Theta_3 \\ \Theta_4, 0 \end{array} \right.\right) dx$$
$$- \frac{\Theta_1}{\Gamma(\tau_1)} \int_0^\infty e^{-bx} x^{a-1} \gamma\left(\tau_1, \frac{x}{\alpha_{1,k}^{(j)} \tilde{\delta}_{SR,j}^2}\right)$$
$$\left. \times G_{\theta+1,3\theta+1}^{3\theta,1}\left(\frac{\Theta_2}{\mu_\theta} x \left| \begin{array}{c} 1, \Theta_3 \\ \Theta_4, 0 \end{array} \right.\right) dx \right]. \quad (43)$$

Further, substituting $\gamma(1+n, x) = \Gamma(n) - \Gamma(n) \exp(-x) \sum_{l=0}^n \frac{x^l}{l!}$ [39, Eq. (8.352.1)] and subsequently employing the integral identities [39, (3.3551.3), (7.813.1)], the above equation can be readily solved to yield the final expression for the floor $\lim_{P_A \to \infty} \overline{P}_e = \frac{1}{JN_bB} \sum_{j=1}^J \sum_{k=1}^{N_b} \sum_{n=1}^B P_e^{(j)}[n,k]$ in (28).

## REFERENCES

[1] Ansari, Imran Shafique and Abdallah, Mohamed M and Alouini, Mohamed-Slim and Qaraqe, Khalid A, "outage performance analysis of underlay cognitive RF and FSO wireless channels," in *proc. 3rd IWOW*, 2014, pp. 6–10.
[2] I. S. Ansari, M. M. Abdallah, M.-S. Alouini, and K. A. Qaraqe, "A performance study of two hop transmission in mixed underlay RF and FSO fading channels," in *proc. IEEE WCNC*, 2014, pp. 388–393.
[3] Y. Deng, M. Elkashlan, P. L. Yeoh, N. Yang, and R. K. Mallik, "Cognitive MIMO relay networks with Generalized selection combining," *IEEE Trans. Wireless Commun.*, vol. 13, no. 9, pp. 4911–4922, 2014.
[4] Y. Deng, L. Wang, M. Elkashlan, K. J. Kim, and T. Q. Duong, "Generalized selection combining for cognitive relay networks over Nakagami-$m$ fading," *IEEE Trans. Signal Process.*, vol. 63, no. 8, pp. 1993–2006, 2015.
[5] D. Tse and P. Viswanath, *Fundamentals of Wireless Communication*. Cambridge University Press, 2005.
[6] E. Dahlman, S. Parkvall, and J. Skold, *4G: LTE/LTE-advanced for mobile broadband*. Academic press, 2013.
[7] N. I. Miridakis, M. Matthaiou, and G. K. Karagiannidis, "Multiuser relaying over mixed RF/FSO links," *IEEE Trans. Commun.*, vol. 62, no. 5, pp. 1634–1645, 2014.
[8] V. Tarokh, H. Jafarkhani, and A. R. Calderbank, "Space-time block codes from orthogonal designs," *IEEE Trans. Inf. Theory*, vol. 45, no. 5, pp. 1456–1467, 1999.
[9] A. Goldsmith, *Wireless Communications*. Cambridge University Press, 2005.
[10] E. Lee, J. Park, D. Han, and G. Yoon, "Performance analysis of the asymmetric dual-hop relay transmission with mixed RF/FSO links," *IEEE Photon. Technol. Lett.*, vol. 23, no. 21, pp. 1642–1644, 2011.
[11] I. S. Ansari, F. Yilmaz, and M.-S. Alouini, "Impact of pointing errors on the performance of mixed RF/FSO dual-hop transmission systems," *IEEE Wireless Commun. Lett.*, vol. 2, no. 3, pp. 351–354, 2013.
[12] E. Zedini, H. Soury, and M.-S. Alouini, "On the performance analysis of dual-hop FSO fixed gain transmission systems," in *proc. ISWCS 15*, August 2015.
[13] H. Samimi and M. Uysal, "End-to-end performance of mixed RF/FSO transmission systems," *IEEE/OSA J. Opt. Commun. and Netw.*, vol. 5, no. 11, pp. 1139–1144, 2013.
[14] E. Zedini, I. S. Ansari, and M.-S. Alouini, "Performance analysis of mixed Nakagami-$m$ and Gamma–Gamma dual-hop FSO transmission systems," *IEEE Photon. J.*, vol. 7, no. 1, pp. 1–20, 2015.
[15] G. T. Djordjevic, M. I. Petkovic, A. M. Cvetkovic, and G. K. Karagiannidis, "Mixed RF/FSO relaying with outdated channel state information," *IEEE J. Sel. Areas Commun.*, vol. 33, no. 9, pp. 1935–1948, 2015.
[16] S. Anees and M. R. Bhatnagar, "Performance evaluation of decode-and-forward dual-hop asymmetric radio frequency-free space optical communication system," *IET Optoelectronics*, vol. 9, no. 5, pp. 232–240, 2015.
[17] ——, "Information theoretic analysis of DF based dual-hop mixed RF-FSO communication systems," in *proc. IEEE PIMRC*, 2015, pp. 600–605.
[18] S. Anees, P. Meerur, and M. R. Bhatnagar, "Performance analysis of a DF based dual hop mixed RF-FSO system with a direct RF link," in *proc. IEEE GlobalSIP*, 2015, pp. 1332–1336.
[19] E. Zedini, H. Soury, and M.-S. Alouini, "On the performance analysis of dual-hop mixed FSO/RF systems," *IEEE Trans. Wireless Commun.*, vol. 15, no. 5, pp. 3679–3689, 2016.
[20] N. Varshney and P. Puri, "Performance analysis of decode-and-forward-based mixed MIMO-RF/FSO cooperative systems with source mobility and imperfect CSI," *J. Lightw. Technol.*, vol. 35, no. 11, pp. 2070–2077, 2017.
[21] H. Arezumand, H. Zamiri-Jafarian, and E. Soleimani-Nasab, "Outage and diversity analysis of underlay cognitive mixed RF-FSO cooperative systems," *IEEE/OSA J. Opt. Commun. Netw.*, vol. 9, no. 10, pp. 909–920, 2017.
[22] F. S. Al-Qahtani, A. H. A. El-Malek, I. S. Ansari, R. M. Radaydeh, and S. A. Zummo, "Outage analysis of mixed underlay cognitive RF MIMO and FSO relaying with interference reduction," *IEEE Photon. J.*, vol. 9, no. 2, pp. 1–22, 2017.
[23] N. Varshney and A. K. Jagannatham, "Cognitive decode-and-forward MIMO-RF/FSO cooperative relay networks," *IEEE Commun. Lett.*, vol. 21, no. 4, pp. 893–896, 2017.
[24] S. M. Kay, "Fundamentals of statistical signal processing, vol. ii: Detection theory," *Signal Processing. Upper Saddle River, NJ: Prentice Hall*, 1998.
[25] W. Cho, S. I. Kim, H. kyun Choi, H. S. Oh, and D. Y. Kwak, "Performance evaluation of V2V/V2I communications: the effect of midamble insertion," in *proc. VITAE*, 2009, pp. 793–797.
[26] G. Acosta-Marum, "Dissertation: Measurement, modeling, and OFDM synchronization for the wideband mobile-to-mobile channel," 2007. [Online]. Available: http://hdl.handle.net/1853/14535\EatDot
[27] G. Acosta-Marum and M. A. Ingram, "Six time and frequency-selective empirical channel models for vehicular wireless LANs," *IEEE Veh. Technol. Mag.*, vol. 2, no. 4, pp. 4–11, 2007.
[28] Y. Khattabi and M. M. Matalgah, "Performance analysis of AF cooperative networks with time-varying links: Error rate and capacity," in *proc. IEEE WTS*, 2014, pp. 1–6.
[29] ——, "Performance analysis of AF cooperative networks with time-varying links: Outage probability," in *proc. IEEE WTS*, 2014, pp. 1–6.
[30] ——, "Conventional AF cooperative protocol under nodes-mobility and imperfect-CSI impacts: Outage probability and shannon capacity," in *proc. IEEE WCNC*, 2015, pp. 13–18.
[31] ——, "Conventional and best-relay-selection cooperative protocols under nodes-mobility and imperfect-CSI impacts: BER performance," in *proc. IEEE WCNC*, 2015, pp. 105–110.
[32] Y. M. Khattabi and M. M. Matalgah, "Performance analysis of multiple-relay AF cooperative systems over Rayleigh time-selective fading channels with imperfect channel estimation," *IEEE Trans. Veh. Technol.*, vol. 65, no. 1, pp. 427–434, 2016.
[33] H. S. Wang and P.-C. Chang, "On verifying the first-order Markovian assumption for a Rayleigh fading channel model," *IEEE Trans. Veh. Technol.*, vol. 45, no. 2, pp. 353–357, 1996.
[34] M. Hanif, H.-C. Yang, and M.-S. Alouini, "Receive antenna selection for underlay cognitive radio with instantaneous interference constraint," *IEEE Signal Process. Lett.*, vol. 22, no. 6, pp. 738–742, 2015.
[35] N. Varshney, A. K. Jagannatham, and P. K. Varshney, "Technical Report: Cognitive MIMO-RF/FSO cooperative relay communication with mobile nodes and imperfect channel state information," IIT Kanpur, Tech. Rep., 2017, [Online] http://www.iitk.ac.in/mwn/documents/MWNLab_TR_CDeco_2017.pdf
[36] G. Bansal, M. J. Hossain, and V. K. Bhargava, "Optimal and suboptimal power allocation schemes for OFDM-based cognitive radio systems," *IEEE Trans. Wireless Commun.*, vol. 7, no. 11, 2008.
[37] A. A. Farid and S. Hranilovic, "Outage capacity optimization for free-space optical links with pointing errors," *J. Lightw. Technol*, vol. 25, pp. 1702–1710, 2007.
[38] M. Niu, J. Cheng, and J. F. Holzman, "Error rate performance comparison of coherent and subcarrier intensity modulated optical wireless communications," *IEEE/OSA J.Opt. Commun. Netw.*, vol. 5, no. 6, pp. 554–564, 2013.
[39] A. Jeffrey and D. Zwillinger, *Table of integrals, series, and products*. Academic Press, 2007.
[40] I. S. Ansari, M.-S. Alouini, and F. Yilmaz, "A unified performance analysis of free-space optical links over Gamma-Gamma turbulence channels with pointing errors," 2013, [Online] http://hdl.handle.net/10754/305353.

[41] C. Zhong, T. Ratnarajah, and K.-K. Wong, "Outage analysis of decode-and-forward cognitive dual-hop systems with the interference constraint in Nakagami-$m$ fading channels," *IEEE Trans. Veh. Technol.*, vol. 60, no. 6, pp. 2875–2879, 2011.

[42] P. J. Smith, P. A. Dmochowski, H. A. Suraweera, and M. Shafi, "The effects of limited channel knowledge on cognitive radio system capacity," *IEEE Trans. Veh. Technol.*, vol. 62, no. 2, pp. 927–933, 2013.

[43] W. Jaafar, T. Ohtsuki, W. Ajib, and D. Haccoun, "Impact of the CSI on the performance of cognitive relay networks with partial relay selection," *IEEE Trans. Veh. Technol.*, vol. 65, no. 2, pp. 673–684, 2016.

[44] N. Varshney and A. K. Jagannatham, "MIMO-STBC based multi-relay cooperative communication over time-selective Rayleigh fading links with imperfect channel estimates," *IEEE Trans. Veh. Technol.*, vol. 66, no. 7, pp. 6009–6025, 2017.

[45] B. K. Chalise and L. Vandendorpe, "Outage probability analysis of a MIMO relay channel with orthogonal space-time block codes," *IEEE Communications Letters*, vol. 12, no. 4, pp. 280–282, 2008.

[46] The Wolfarm Functions Site. [Online]. Available: http:/functions.wolfarm.com

[47] J. M. Romero-Jerez and A. J. Goldsmith, "Performance of multichannel reception with transmit antenna selection in arbitrarily distributed Nakagami fading channels," *IEEE Trans. Wireless Commun.*, vol. 8, no. 4, pp. 2006–2013, 2009.

[48] I. S. Ansari, S. Al-Ahmadi, F. Yilmaz, M.-S. Alouini, and H. Yanikomeroglu, "A new formula for the BER of binary modulations with dual-branch selection over Generalized-K composite fading channels," *IEEE Trans. Commun.*, vol. 59, no. 10, pp. 2654–2658, 2011.

[49] H. Chergui, M. Benjillali, and S. Saoudi, "Performance analysis of project-and-forward relaying in mixed MIMO-pinhole and Rayleigh dual-hop channel," *IEEE Commun. Lett.*, vol. 20, no. 3, pp. 610–613, 2016.

[50] M. Shah, "On generalizations of some results and their applications," *Collectanea Mathematica*, vol. 24, no. 3, pp. 249–266, 1973.

[51] B. Sharma and R. Abiodun, "Generating function for generalized function of two variables," *Proceedings of the American Mathematical Society*, vol. 46, no. 1, pp. 69–72, 1974.

[52] J. Lee, H. Wang, J. G. Andrews, and D. Hong, "Outage probability of cognitive relay networks with interference constraints," *IEEE Trans. Wireless Commun.*, vol. 10, no. 2, pp. 390–395, 2011.

[53] H. Ding, J. Ge, D. B. da Costa, and Z. Jiang, "Asymptotic analysis of cooperative diversity systems with relay selection in a spectrum-sharing scenario," *IEEE Trans. Veh. Technol.*, vol. 60, no. 2, pp. 457–472, 2011.

[54] A. K. Meshram, D. S. Gurjar, and P. K. Upadhyay, "Joint impact of nodes-mobility and channel estimation error on the performance of two-way relay systems," *Physical Commun.*, vol. 23, pp. 103–113, 2017.

[55] A. P. Prudnikov, Y. A. Brychkov, and O. I. Marichev, *Integrals and Series Volume 3: More Special Functions*, 1st ed. Gordon and Breach Science, 1986.